\documentclass[fleqn,usenatbib]{mnras}

\usepackage{newtxtext,newtxmath}

\usepackage{sidecap}

\usepackage[T1]{fontenc}

\DeclareRobustCommand{\VAN}[3]{#2}
\let\VANthebibliography\thebibliography
\def\thebibliography{\DeclareRobustCommand{\VAN}[3]{##3}\VANthebibliography}

\usepackage[normalem]{ulem}

\usepackage{graphicx, caption, subcaption}	
\usepackage{amsmath}	

\usepackage[dvipsnames]{xcolor}

\title[]{The influence of galaxy mergers, black-hole growth, and gas processes on the evolution of the stellar mass-gas metallicity relation of galaxies in different cosmic environments}

\author[A.R. Rowntree et al.]{Aaron R. Rowntree$^{1}$ \thanks{E-mail: a.rowntree-2018@hull.ac.uk (ARR)}, Fiorenzo Vincenzo$^{2}$, Ankit Singh$^{3,4}$, Gareth Few$^{1}$, Jaehyun Lee$^{5}$, \newauthor Christophe Pichon$^{3,6,7}$
\\ ~ \\
$^{1}$E.~A. Milne Centre for Astrophysics, University of Hull, Hull, HU6 7RX, UK\\
$^{2}$Dipartimento di Fisica e Astronomia “Ettore Majorana”, Università degli Studi di Catania, Via S. Sofia 64, 95123 Catania, Italy\\
$^{3}$Korea Institute for Advanced Study (KIAS), 85 Hoegiro, Dongdaemun-gu, Seoul 02455, Republic of Korea\\
$^{4}$ School of Physics and Astronomy, University of Nottingham, University Park, Nottingham NG7 2RD, United Kingdom\\
$^{5}$Korea Astronomy and Space Science Institute, 776, Daedeokdae-ro, Yuseong-gu, Daejeon 34055, Republic of Korea\\
$^{6}$Institut d’Astrophysique de Paris, UMR 7095, CNRS, Sorbonne Université, 98 bis boulevard Arago, 75014 Paris, France\\
$^{7}$Kyung Hee University, Dept. of Astronomy and Space Science, Yongin-shi, Gyeonggi-do 17104, Republic of Korea\\
}

\date{Accepted 202- xxx; Received 202- xxx; in original form 202- xxx}
\pubyear{202-}

\begin{document}

\maketitle

\begin{abstract}

We study the impact of supermassive black hole (SMBH) growth, $\langle \dot{M}_\mathrm{SMBH}\rangle$, major and minor galaxy mergers, and gas processes, on the average gas metallicity of galaxies within the Horizon Run 5 simulation, with the aim to uncover which of these processes drive the scatter in the gas metallicity-stellar mass relation (MZR) at different redshifts in nodes, filaments and voids. At $z=5$, minor mergers produce the largest differential in $\log[Z_g/Z_\odot]$ where all environments display a maximum $0.22$ dex increase in the average $\log[Z_g/Z_\odot]$ compared to non-merging galaxies. The node population also displays a consistent $0.1$ dex negative offset in $\mathrm{d} \log[Z_g/Z_{\odot}]$, the residual $Z_g$ from the total MZR of all galaxies, across all redshifts, whilst filament and void galaxies show a smaller offset. Major mergers show little influence on these same properties. This suggests minor mergers regulate metallicity and contribute to galaxy mass growth concurrently, accelerating chemical evolution post merger. Between $z=1-3$, a high $\langle \dot{M}_\mathrm{SMBH}\rangle$ leads to a larger negative offset in $\mathrm{d} \log[Z_g/Z_{\odot}]$ for all environments. Here, node galaxies show the largest negative offset of approximately $0.25$ dex, suggesting that AGN-driven gas removal may contribute to the MZR scatter at intermediate times. Finally, galaxies with low $M_{gas}/{M_{tot}}$ show increased $\mathrm{d} \log[Z_g/Z_{\odot}]$ across all redshifts and environments, again a $0.25$ dex maximum for node galaxies. These galaxies also spike in $\mathrm{d} \log[Z_g/Z_{\odot}]$ at late times, below $z=1$. At this time, galaxies in the nodes show negative $\langle \dot{M}_\mathrm{gas} \rangle$ whilst also showing the largest $\mathrm{d} \log[Z_g/Z_{\odot}]$ values we observe of $0.2$ dex, suggesting the importance of the balance between gas accretion and starvation in driving MZR scatter at low redshifts.

\end{abstract}


\section{Introduction}
\label{sec:intro}

Galaxies evolve as a function of the processes that act upon them: e.g. galaxy mergers \citep{Toomre1972, LarsonTinsley1978, Ellison2022, Sparre2022, Duan2025}, gas inflows and outflows \citep{Schmidt1959, ReesOstriker1977, Tacconi2020}, and SMBH growth, or AGN activity, \citep{Seyfert1943, Salpeter1964, Silk1998, Cresci2023}. Interestingly, the efficiency, frequency, and impacts of these physical processes varies depending on a galaxy's surrounding environment \citep{Dressler1980}: mergers \citep{2012Jian, Sureshkumar2024}, AGN activity \citep{Kirshnan2017, Malavasi2022, Mountrichas2023, DelPino2023}, and gas dynamics \citep{Zabel2019, Song2021}. This means that a galaxy's evolution is fully dependent on its own unique path through physical space and time itself. 

The Universe is populated with discrete cosmic environments that have significantly different densities. This density field makes up the cosmic web \citep{Bond1996}, a striking pattern seen in the matter distribution of the Universe. The cosmic web was seeded in the pre-inflationary era by quantum over-densities in the dark matter (DM) density field. These density contrasts were amplified due to rapid inflation, creating significant gravitational potentials that grew over time \citep{Peebles1969, Zeldovich1970}. At present this formation of structure has led to unique environments that galaxies reside in. Three commonly studied environments are nodes, the highest galaxy-density regions associated with the aforementioned large gravitational potentials \citep{Gregory1978}, filaments, the intermediate density pathways connecting the nodes \citep{Zeldovich1970, deLapperant1986, Gregory1978} and the voids, the almost empty regions of low density, \citep{Kirshner1987}.

Within these different densities, physical processes act at different efficiencies due to the proximity of galaxies to one another, and the density of their surrounding IGM, depending on the process or feedback mechanism in question \citep{GunnGott1972, Lin2010, Maier2019, Mountrichas2023, Hoosain2024}. Therefore, galaxies show differences in their properties depending on the their environment. In both simulations and observations, SFR is lower in higher density regions where starvation and quenching mechanisms are more efficient \citep{Peng2010, Haines2011, 2016Alpaslan, 2016Mart, Kraljic2018, Mahajan2018, 2020Singh, Gallazzi2021, Malavasi2022, Hasan2023}.  Stellar mass increases with higher densities{\citep{Vulcani2012, Cucciati2017}, showing the importance of environment as a key driver of galaxy evolution \citep{Peng2010, 2016Alpaslan, Rowntree2024}. Gas fractions are seen to be significantly higher for galaxies in the field, or void, where chemical evolution and gas consumption is not accelerated significantly by feedback processes and availability of cold gas is far higher \citep{Pustilnik2016, Kreckel2016, Rowntree2026}. Finally, metallicity correlates positively with density, strongly linking to enriched inflows in nodes, shorter quenching timescales in nodes, and again, the higher availability of cold gas to isolated galaxies \citep{Shields1991, Henry1992, Cooper2008, Peng2014, Donnan2022, Rowntree2024}.

In higher density regions, a higher frequency of galaxy-galaxy interactions, specifically mergers, are expected \citep{Lin2010, Kocevski2011, deRavel2011, 2012Jian, Lotz2013, Kampczyk2013, Watson2019, Pearson2024, Shojaei2025}, due to lower nearest neighbour distances \citep{Liu2023, Laishram2024, Shibyua2025}. However, galaxy clusters display high velocity dispersions that act to reduce merger rates \citep{Sureshkumar2024}, making the relationship between merger rate and environment a balance between these properties.
The merger mass ratio defines both minor (both masses are significantly different) and major (where both have similar mass) mergers \citep{Cox2008, Lotz2010, Lopez2012}. Major mergers typically increase SFR in the involved galaxies \citep{Knapen2015}; particularly in gas-rich mergers \citep{Lambas2012, Knapen2015, Kaviraj2015}, emerging from gas-rich inflows triggered by the tidal torque in the merger \citep{Lambas2012}. However, \cite{Pearson2019} showed that mergers may not be as important for driving SFR as other processes, particularly at early times \citep{Kaviraj2015}. \citet{Lambas2012} suggests that, as much as major mergers are the most destructive and impactful merger type, it is encounters between galaxies and minor companions that account for most $M_\star$ evolution.

We also study the evolution of the average supermassive black hole (SMBH) mass in time. This quantity has been shown to directly relate to AGN activity, \citep{Soltan1982, Marconi2004, DiMatteo2005, Booth2009}. Over the last few decades, AGN have been shown to both negatively \citep{Bower2006, BrandtAlexander2015, Santoro2020, Contini2024} and positively \citep{Venturi2023, Gim2024, Das2025} impact the SFR of galaxies. Higher density environments boast higher AGN-fraction between $z\approx1-3$ \citep{Gatica2024}, and a general trend that the physical number of AGN increases with environmental density exists \citep{Hwang2012}. Around the higher density environments, nodes and filaments, the impact on SFR is more prominent \citep{Kraljic2019}. As such one could expect differences in galaxy properties based on AGN activity between environments.

In our previous work \citep{Rowntree2024, Rowntree2026}, we focus on the effect that environment has on gas metallicity and the Gas Metallicity-Stellar Mass relation (MZR). Understanding this relation and how it arises is key in understanding the physical processes that influence the galaxies themselves \citep{Tremonti2004}. The MZR has been studied at low redshift \citep{1979Lequeux,Tremonti2004}, higher redshift \citep{2005Savaglio, 2008Maiolino, 2013Zahid}, and more recently across much larger redshift ranges in the MOSDEF \citep{Sanders2021} and JADES, CEERS and UNCOVER surveys \citep{Pallottini2025}. Large-scale cosmological simulations like IllustrisTNG \citep{Nelson2019}, Horizon Run 5 \citep{Lee2020} and SIMBA \citep{Dave2019}, have also provided theoretical predictions into areas that current surveys cannot. The MZR shows a significant scatter across the full $M_\mathrm{\star}$ range \citep{Tremonti2004, 2005Savaglio, Ellison2008}, where the scatter is particularly large for low $M_\mathrm{\star}$ and low redshift. This scatter has been directly tied to SFR and the three parameters together constitute what is known as the fundamental metallicity relation (FMR) \citep{Mannucci2010, LaraLopez2010, Cresci2012, Yates2012, HaydenPawson2022}. Work in \cite{Bothwell2013} and \cite{Torrey2019} instead links the MZR scatter directly with gas mass, with \cite{DeRossi2017} and \cite{vanLoon2021} extending this further to consider gas fraction, finding the strongest dependencies on metallicity for both properties. However, recent studies in \cite{Menguiano2024} and \citet{Koller2025} report that baryonic gravitational potential outweighs SFR and even stellar mass in significance. Environment itself has frequently been studied relative to the MZR, establishing a strong connection between the two, \citep{Shields1991, Henry1992, Skillman1996, Cooper2008, 2009Ellison, 2017Wu, Gupta2018, Donnan2022, Andrade2024, Kane2024}, where environment directly links into all aforementioned galaxy properties through the processes at play. The MZR also has an evolution in time. Its normalisation systematically increases towards lower redshift, whilst its slope remains relatively invariant \citep{2008Maiolino, Zahid2014, Torrey2019, Sanders2021, Langeroodi2023}. 

It is really the gas in galaxies that defines their potential to form stars and to continue to evolve. Gas consumption timescales are far shorter than that of the age of the Universe, as such, galaxies must continually acquire gas to form stars \citep{Kacprzak2017}. Its accepted that cold accretion plays a main role in the growth of galaxies \citep{ReesOstriker1977}, particularly for lower mass halos \citep{Keres2005}, acting as the main driver of the cosmic star formation history \citep{Keres2009}. Results from \cite{Cresci2010} suggested that at $z=3$ this cold accretion reduces the metallicity of regions that it is funnelled into, hinting that chemical dilution could be a signature of cold gas accretion. \citet{Sanders2015} further found, at $z=2.3$ in the MOSDEF survey, that their sample of star forming galaxies were offset from the FMR, suggesting a phase of galaxy growth where cold inflow rates exceed that of star formation and outflows. 
Gas accretion rates, and the availability of cold gas, differs between environment. Galaxies in nodes lose the gas in their reservoirs at increased rates, leading to earlier onset of starvation and quenching \citep{VandeVoort2017}. This leads to reduced SFR and rapid increase in metallicity without regulation from cold gas accretion \citep{Peng2015, Baker2024}. Void galaxies, by contrast, are regions that are less dense both in terms galaxy number and total mass. At fixed (sub)halo mass their haloes assemble later and accrete gas less efficiently than their counterparts in filaments and nodes, so their baryonic and chemical evolution is delayed rather than enhanced \citep{RodriguezMedrano2021, RosasGuevara2022}. This slower processing of gas into stars and metals leaves void galaxies comparatively gas-rich and metal-poor down to low redshift.

In this paper we study how these three physical processes, galaxy mergers, AGN feedback and gas processes impact the metallicity of galaxies in the three environments across a redshift range of $z=0.625$ to $5$. We use the {\tt T-ReX} filament finding algorithm to define the cosmic structure in the HR5 simulation. Using this, we define the nodes, filaments and voids in our dataset. We aim to uncover which of the processes is the dominant driver of gas metallicity and the MZR in each environment at different epochs to better understand how a galaxy becomes chemically enriched. By defining the MZR across time in the three environments, we can also determine whether these effects are creating galaxies above or below the population's average $Z_g$ separate to the known trends with $M_\star$.

\section{Method}
\label{sec:Method}

For this study we use the Horizon Run 5 (HR5) cosmological hydrodynamical simulation \citep{Lee2021}. This simulation models the evolution of the Universe down to z=0.625, tracking Mpc to kpc scales made possible by the utilization of an MPI-OpenMP version of the adaptive mesh refinement (AMR) code RAMSES \citep{Teyssier2002}. The box size of HR5 is $1.049 \times 1.049 \times 1.049 \ \text{cGpc}^{3}$, but within this a smaller cuboid region of $1.049 \times 0.119 \times 0.127 \ \text{cGpc}^{3}$ is set as the high resolution zone, within which grid cells can be refined down to $\Delta x = 1 \ \text{kpc}$, depending on the local density of the cell. The simulation employs a range of sub-grid physical processes: supernovae (SN) \citep{Dubois2008} and AGN feedback \citep{Dubois2012}, star-formation activity \citep{Rasera2006}, ultraviolet background heating \citep{Haardt1996} and metallicity-dependent radiative cooling \citep{Dalgarno1972}. Galaxy merger trees are constructed based on stellar particle membership in galaxies \citep{Lee2024}, following its properties with time as mergers happen. A galaxies chemical evolution in HR5 is governed by the results of RAMSES-CH \citep{Few2012}, which includes a continuous feedback model. This allows for energy and chemical feedback to be followed through the AMR grid, providing constraints on key parameters like abundance ratios and supernovae rates. This model includes both core-collapse and Type Ia supernovae, providing the means to track the evolution of Oxygen and Iron abundances in both the ISM and galaxies' stellar populations.
Note that the specific redshifts and mass scales mentioned in this study, with reference to HR5, should only be interpreted qualitatively as they are sensitive to the simulation parameters. HR5 follows the $\Lambda$ cold dark-matter ($\Lambda CDM$) Universe with the following cosmological parameters: $\Omega_{0}=0.3$, $\Omega_{\Lambda}=0.7$, $\Omega_{\rm b} =0.047$, $\sigma_{8} = 0.816$, and $H_{0}= 100\times h_{0}=68.4\,\text{km}\,\text{s}^{-1}\,\text{Mpc}^{-1}$, which are compatible with the Planck data \citep{Plank2016}.

\subsection{Galaxy selection}
\label{subsec:GalSel}

HR5 outputs contain galaxy and halo catalogues at 128 snapshots between $z=15.555$ and $z=0.625$. The catalogues contain all key properties that describe galaxies and large halos (galaxy clusters). They also include important flags for tracking galaxy merger history. The galaxy and halo catalogues are populated using an extended friends-of-friends (FoF) algorithm that identifies virialized halos through a chain of linkages between DM, stellar and black hole (BH) particles. Groups of galaxies are then handed to the PSB-Based Galaxy Finder ({\tt PGalF}), which is based on the Physically Self-Bound (PSB) algorithm \citep{Kim2006}. {\tt PGalF} identifies subhalos (galaxies) based on peaks in the $M_\star$ density field. Each peak has possible core and non-core particles, which are extracted creating a galaxy candidate. A membership decision for each particle is then ran, checking whether it lies within the tidal boundary of the galaxy, combined with ensuring the total energy of the particles within the boundary returns a valid result. This galaxy is then added to the subhalo catalog, refer to \citet{Lee2021} and \citet{Kim2023} for more details.

This work selects the halo and subhalo catalogues at 86 snapshots between $z=0.625$ and $z=4.5$. Each halo catalogue contains the FoF groups of galaxies at each snapshot, and each subhalo catalogue contains the galaxies at each snapshot. To ensure the galaxies we use in this study are valid, and resolved well enough to contain sufficiently accurate galaxy properties, we implement one main cut that applies at all redshifts, and one cut that applies only at $z=0.625$. The first cut that applies equally to all snapshots is one that follows the {\tt pure} flag in the galaxy catalog. This flag identifies galaxies that sit too close to low-resolution DM particles within the simulation. These galaxies are removed to ensure that the low-resolution particles do not impact galaxy properties in unwanted ways. The second cut that is only carried out at $z=0.625$ is a $M_\star$ cut. We take the $M_\star$ cut at $M_{\star} > 2 \times 10^9 M_{\odot}$. A stellar particle in HR5 has a minimum mass of $2\times10^6 M_\odot$, and as such, after this cut, all galaxies at $z=0.625$ have at least 1000 stellar particles, ensuring their resolution is high enough for the computation of accurate galaxy properties. These two cuts at $z=0.625$ give us $158,094$ galaxies to follow back in time to $z=5$.

In this work we study stellar mass, $M_\star$, gas mass, $M_\mathrm{gas}$, gas metallicity, $Z_g$ and SMBH mass, $M_\mathrm{SMBH}$. In HR5, gravitational boundedness is computed for every matter component. This means that the stars, DM and gas that belong to a galaxy are defined as those that are gravitationally bound to that galaxy. In the case of gas, there is no explicit classification into ISM and circum-galactic medium (CGM) components. Throughout this work, all masses refer only to material that is gravitationally bound to individual galaxies; diffuse, unbound baryons are not included in any of our environmental budgets. We also make use of certain flags in the catalogue to compute the merger tree of the selected galaxies. To collect the data needed for this study, we follow each galaxy back to its main progenitor in the previous snapshot, making note of each of the relevant galaxy properties at each snapshot. This provides us with the history of each property and how it changes between each snapshot. We used these values along with the change in time between snapshots to compute the time-averaged change in gas mass, $\langle \dot{M}_\mathrm{gas} \rangle$, and the time-averaged SMBH growth rate, $\langle \dot{M}_\mathrm{SMBH} \rangle$. These quantities are calculated as seen in Eq. \ref{eq:TimeAveraged}.

\begin{equation}
\label{eq:TimeAveraged}
\langle \dot{M}_{\rm X} \rangle \equiv \frac{M_{\rm X}(z_2)-M_{\rm X}(z_1)}{t(z_2)-t(z_1)}
\end{equation}

Where X refers to the specific type of mass we refer to, and the numbers 1 and 2 refer to the snapshot. Finally, for certain properties we choose to report the fractional quantity rather than the absolute quantity to eliminate any effects from the secular evolution of the galaxy. Throughout the work, $M_\mathrm{gas}$, $M_\mathrm{\star}$ and the respective change in these properties are reported relative to the total mass of the galaxy $M_{tot} = M_{gas} + M_{\star}$.

It is important to note that the time between snapshots is not consistent in HR5. There is typically a small variance around a systematically increasing trend in $\Delta t$ with decreasing redshift. However, certain snapshots display a dramatically lower $\Delta t$ than the expected, which also leads to non-physical variance in the trends of other galaxy properties, only within these snapshots. These snapshots have been manually removed from the histories of each galaxy, leaving only snapshots that have a $\Delta t$ within the expected variance such that we see a more consistent $\Delta t$ across the full range.
This method leaves us with a way to track each galaxy's evolution in the relevant properties between $z=0.625$ and $z=5$, a way to track whether a merger is, or is not, occurring, and a way to measure the AGN activity of a galaxy across time. What follows is how we then calculate which unique environment each galaxy exists within.

\subsection{Structure Estimation}

To determine the environment that a galaxy resides in, we must be able to quantify the environments themselves. This is typically achieved through the use of a filament finding algorithm, such as {\tt DisPerSE} \citep{Sousbie2011}, {\tt Nexus} \citep{Cautun2013} and {\tt T-ReX} \citep{Bonnaire2020, Bonnaire2022}. For a comprehensive comparison of different filament finders, their methodologies and their outputs, refer to \citep{Libeskind2018}.
In this study, like our prior studies \citep{Rowntree2024, Rowntree2026}, we use {\tt T-ReX} as our filament finding algorithm. In short, {\tt T-ReX} is purpose-built to be used on a distribution of galaxies. {\tt T-ReX} specifically first takes an input of a galaxy distribution. From this, Gaussian mixture models (GMM) are used to describe the distribution of the tracers and graph theory, more specifically a minimum spanning tree, is used to produce a smooth connection between the tracers \citep{Bonnaire2020}. The outputs from the algorithm are a list of edges in Cartesian space that combine to create a ``skeleton" that estimates the large-scale structure present in the distribution of the galaxies. Using this filament finder and our method allows us to be closely comparable to future observational studies, without sacrificing the accuracy and robustness of structure estimates. For further detail on {\tt T-ReX} and surrounding considerations, refer to both \citet{Rowntree2024} and \citet{Rowntree2026}. 
In this study we make use of the structure estimates produced in our previous paper. These are constituted of 9 skeletons produced for datasets at 9 snapshots between $z=0.625$ and $z=5$ in increments of $0.5$. To ensure the comparability of the structure estimates, and properties that depend on them, we fix the galaxy density to $\rho_i=0.02519$ in each dataset, and use the same {\tt T-ReX} parameter set of $\lambda = 30$, $\sigma = 0.38$ and $l=25$ in each computation. For detail on the reasoning behind these steps, and the parameter tuning, read the methodology section in \cite{Rowntree2026}. A short overview of the parameters is that $l$, is a de-noising parameter that defines the level in the tree at which all branches beyond are removed. $\sigma$ defines the extent of the gaussian clusters; a small $\sigma$ leads to a more coarse structure, whilst a large $\sigma$ produces a smooth structure. $\lambda$ is the elastic constraint of the tree, the larger it is, the shorter and smoother the tree is. This methodology produces the 9 skeletons that we use to define our three unique cosmic environments.

\subsection{Environmental Definitions}

Using these 9 skeletons, we have now quantified the large-scale structure in HR5, and we can use this to define the environments we are interested in. We focus on 3 main environments, nodes, filaments and voids. This section will overview how we define each environment.

\subsubsection{Nodes}

To define this environment we use the largest galaxy clusters at each snapshot as a proxy. It is likely that these galaxy clusters lie at the intersections of filaments, as such, they are very commonly the nodes of the skeleton. To select the clusters we use as the proxy in each snapshot, we first take a halo mass cut of $M_{tot} \geq 10^{13}$ at $z\sim 0.625$. This cluster cut defines a certain percentile of the total population of clusters. We then use this percentile in the other snapshots to select the most massive clusters at that time. This accounts for the expected evolution of cluster mass with redshift that using a static mass threshold would not. Note that in this study, as we are only focussed on the environmental definitions of galaxies at $z=0.625$, we only use the cluster definitions at the latest snapshot to define our node galaxies. The broader methodology here is included for completeness. 
The final cut we take in the clusters is to ensure we only look at the relaxed or virialized clusters. By calculating the distance between each cluster's center of mass and the position of the brightest central galaxy, $d_{BCG}$, normalising it to each cluster's $R_{200}$, we define the normalised offset, $\Delta r$. A large offset typically represents an unrelaxed system, so we limit our selection to 
$\Delta r < 0.05$. This selection gives us the halos, galaxy clusters, that we class as nodes. By cross-referencing each galaxy catalogue with these halos, we can then identify which galaxies lie within $2 \times R_{200}$ or the cluster center, giving us our population of node galaxies. Refer to \citet{Rowntree2026} for further specifics on this process.

\subsubsection{Filaments}

We define filament galaxies by using the structure estimates, skeletons, from {\tt T-ReX}. The $d_\mathrm{skel}$ values calculated earlier, are again used here to provide a quantity to identify which galaxies lie close to the skeleton.$d_\mathrm{skel}$ defines the perpendicular distance from a galaxy to its nearest edge, or filament, on the skeleton. This is computed using the ${\tt radial\_distance\_skeleton}$ function within the {\tt T-ReX} library. Note that we only calculate $d_\mathrm{skel}$ and carry out this process for galaxies that lie outside of $2\times R_{200}$ of any previously identified node, as they already belong to another environment. A galaxy that has $d_{skel} < 1$ cMpc is defined to be part of the filament galaxy population. The choice to use $1$ cMpc as our filament boundary is inspired by other work in the field that shows the typical filament radii to be $1-3$ Mpc \citep{Gheller2015, Galarraga2022, Wang2024}. Setting the boundary at the lower end of this range imposes the strictest constraint on filament membership and, as such, provides a population of galaxies that we are the most confident in. Although filament radii evolve with redshift, we approach the definition of filament galaxies the same way in every snapshot, using $d_{skel} < 1$ cMpc to define the population. In the appendix of \citet{Rowntree2026}, we implemented an evolving filament membership boundary, and found that this had no effect on the results.

\subsubsection{Voids}

The void galaxies are also calculated based on $d_\mathrm{skel}$ but instead we look at galaxies that are far from the skeleton instead of nearby. We choose a void boundary of $d_{skel} > 8$ cMpc for which galaxies above this value are considered void galaxies. This cut follows the initial definition from \citep{Rowntree2024}, inspired by the total distribution of $d_\mathrm{skel}$. For more detail on this refer to \cite{Rowntree2024}. Again, this definition remains consistent across redshift.

Note that between the filament definition of $d_{skel} < 1 \ \mathrm{cMpc}$ and the void definition $d_{skel} > 8 \ \mathrm{cMpc}$ lies a region of unused data points. The decision to have the boundary of these two environments so distinct from one another is to remove the "fuzzy" region between the two, and to increase confidence that the galaxies we extract from these definitions lie within our the regions we aim to probe. It is difficult to know the true boundary of the start of the filament, or the end of the void, so it is more simple to push our filament environment boundary closer to the filament spine, or to push our void boundary closer to the void center. This region between the two accounts for approximately 10\% of galaxies in each snapshot, and as such we are still left with sufficiently large populations of galaxies in each environment for analysis.

\subsection{Physical Processes}

The following section will outline how we track the influence of each of the physical processes that we aim to study. These processes are SMBH growth, galaxy mergers and gas accretion/change in gas mass. Each of these processes needs to be quantifiable within this study to see its impact.

\subsubsection{Galaxy Mergers}

To define a galaxy merger in HR5 we look to the subhalo catalog. Each snapshot in this catalog provides all the galaxies, and how they progress forward or backwards in time. A single galaxy in this catalogue will point to its main progenitor in the previous snapshot, and if a merger occurred, it will also point to the other progenitors that were involved in the merger. To reconstruct the merger tree of the galaxies in HR5, the catalogues provide a handful of flags that denote a few key things: the link between a galaxy and its progenitors one snapshot prior, and whether a merger has occurred or not. We compute the merger tree for every galaxy selected by our cuts in the snapshot at $z=0.625$, back to $z=5$. As we populate the merger tree, we also take note of the key properties we wish to study, stellar mass, $M_\star$, gas mass, $M_\mathrm{gas}$, gas metallicity, $Z_g$ and BH mass, $M_\mathrm{SMBH}$, providing us with each galaxy's history of every property. At each point in a galaxy's history where there are multiple progenitors, we state that a merger occurs. We calculate whether the merger was minor or major using the mass ratio between the main galaxy, and the next largest galaxy involved in the merger. The mass used to calculate this ratio is $M_{tot}$, defined earlier as $M_{tot} = M_{gas} + M_{\star}$. If this mass ratio is above $0.3$, then the merger is considered a major merger; if it is lower than $0.3$, then the merger is considered a minor merger. This is implemented as inspired by typical methodology found in the surrounding field, and altered to fit our specific dataset \citep{Lotz2010, Lotz2011, LopezSanjuan2012, Lambas2012, Duncan2019, Conselice2022}. To then define a merger rate at each snapshot for each environment, we take the population of galaxies at a specific time, in a specific environment, and sum how many of them are currently undergoing a merger. We then calculate the percentage of the pre-chosen population that this represents, giving us the percentage of galaxies at this time in the nodes, filaments or voids, that are merging. Using this we then have access to three main populations of galaxies in the context of mergers. Those undergoing a major merger, those undergoing a minor merger, and those that are not merging. Combining these definitions with trends in metallicity, $\log[Z_{g}]$, the residual from the MZR of all galaxies, $\mathrm{d} \log[Z_g/Z_{\odot}]$, and environment, we can see the part that mergers play in the chemical evolution of galaxies across time.

\subsubsection{Supermassive Black Hole Growth Factor}

To gain some information regarding the SMBH activity of the galaxies in our samples, we turn to a proxy which is the time-averaged supermassive black hole growth factor, $\langle \dot{M}_\mathrm{SMBH} \rangle$, \citep{Soltan1982, Marconi2004, DiMatteo2005, Booth2009}. To acquire this growth factor for each galaxy, we follow the equation laid out in section \ref{subsec:GalSel}. HR5 models massive black hole (MBH) formation, growth, and AGN feedback following \citet{dubois21}. BHs are seeded with initial masses of $10^4 \ \mathrm{M_\odot}$ in gas cells exceeding the density thresholds of $n_{H}=0.1 \ {\rm cm^{-3}}$, the number density of Hydrogen. To avoid multiple seeds in a single galaxy, a new BH is created only if no pre-existing BH lies within 50 kpc. In HR5, unresolved gas dynamical friction is included following \citet{Dubois13drag}, with a density-dependent boost factor $\alpha_{\rm gas}=\max[(\rho/\rho_0)^2,1]$, where $\rho$ is the gas density within a cell, and $\rho_0$ is the threshold density for star formation. BHs grow through Bondi-Hoyle-Lyttleton accretion, limited by the Eddington rate, with kernel-weighted local gas properties used to estimate the accretion flow \citep{Dubois12}. Once massive enough, their accretion is largely enhanced since the Bondi accretion rate is proportional to the square of the BH mass. At the same time, this produces and releases a large amount of energy. In this regime, BHs begin to self-regulate their growth. In their analysis of HR5, \citet{Lee2021} also noted that the quasar-mode feedback, coupled with the on-the-fly evolution of black-hole spin, slows black-hole growth at high redshift. An analogous time evolution of $M_\mathrm{SMBH}$ is seen in cosmological simulations built on similar physics \citep{Bower2017, Lapiner2021}, and it sets in earlier in the denser node and filament environments where the requisite black-hole masses are reached first.

BH pairs are merged when their separation falls below $4\Delta x$, where $\Delta x$ is the minimum cell size, $\approx 1 \ \rm{kpc}$, and their relative velocity is smaller than the binary escape velocity. AGN feedback operates in two modes switched by the Eddington ratio, $\chi=\dot M_{\rm BH}/\dot M_{\rm Edd}$: a thermal (quasar) mode for $\chi>0.01$ and a jet (radio) mode for $\chi\le0.01$ \citep{Dubois12}, where $\dot{M}_{BH}$ is the BH growth rate and $\dot{M}_{Edd}$ is the Eddington accretion rate. In the thermal mode, feedback injects energy at a rate $\dot E_{\rm BH,h}=\epsilon_{\rm r}\epsilon_{\rm f,h}\dot M_{\rm BHL}c^2$ with $\epsilon_{\rm f,h}=0.15$, where $\dot M_{\rm BHL}$ is the Bondi-Hoyle Lyttleton accretion rate. In the jet mode, the feedback power is $\dot E_{\rm BH,j}=\epsilon_{\rm f,j}\dot M_{\rm BHL}c^2$, where $\epsilon_{\rm f,j}$ depends on BH spin following the magnetically choked accretion flow model of \citet{mckinney12}. BH spins evolve self-consistently through gas accretion and BH mergers \citep{Rezzolla08,Dubois14spin}. Jets deposit mass, momentum, and energy into bipolar outflows with a mass loading factor of $\eta=100$. We note that this $\eta=100$ mass loading applies to the low-accretion radio (jet) mode, which dominates in massive haloes at late times; at the intermediate redshifts ($z \sim 1$ to $3$) where our high-$\langle \dot{M}_\mathrm{SMBH} \rangle$ galaxies accrete close to the Eddington limit, feedback is instead dominated by the thermal quasar mode \citep{Dubois12}, which heats and removes the dense, metal-enriched gas from the centres of massive hosts.

In HR5, $M_\mathrm{SMBH}$ is calculated as the mass of the largest individual SMBH particle in a sub-halo. This means that in the case of a merger, it is possible that the SMBH mass for the galaxy pre and post event is altered. This would appear as SMBH growth, whilst the individual BH has not seen significant accretion, and as such, no relevant BH growth. To account for this, prior to the merger, we sum the $M_\mathrm{SMBH}$ values of all progenitors and compare this value to the post-merger $M_\mathrm{SMBH}$ value for the main galaxy. This means that post-major merger, $\langle \dot{M}_\mathrm{SMBH} \rangle$ can appear negative, and these values can be removed from the dataset. This is an uncommon occurrence in our data, and the removal of these negative values showed no significant impact on our results. In this property, galaxies with high $\langle \dot{M}_\mathrm{SMBH} \rangle$ are the galaxies are experiencing rapid SMBH growth, whilst those with low $\langle \dot{M}_\mathrm{SMBH} \rangle$ are those that are dormant or experiencing very little accretion. To define galaxies with high and low $\langle \dot{M}_\mathrm{SMBH} \rangle$ we take an upper, 95, and lower, 5, percentile cut in the property at each snapshot to create the selection.

\subsubsection{Gas Accretion or Change in Gas Mass}

We calculate the change in gas mass, $\langle \dot{M}_\mathrm{gas} \rangle$, similarly to $\langle \dot{M}_\mathrm{SMBH} \rangle$, by following Eq. \ref{eq:TimeAveraged}. in section \ref{subsec:GalSel}. For this property we do not compute the sum of $M_\mathrm{gas}$ for all progenitors in the case of a merger, as we are interested in the total $M_\mathrm{gas}$ available to the galaxy. What we do infact change is that this property is fractional relative to $M_\mathrm{tot}$. We then consider the variations in the gas fraction, that we define as the ratio of $M_{\text{gas}}$ to $M_{\text{tot}}$; this way the changes in the gas mass for galaxies within a given environment can be better compared on the same scale to one another, irrelevant of galaxy's total mass. We note that this is not a clear measurement of gas accretion, as the gas mass of a galaxy is influenced by many other processes. From this property, we only aim to gain insight into how the total change in the gas reservoir of galaxies influences its chemical evolution. By creating two populations, galaxies that are gaining gas rapidly, and those that are losing gas rapidly, we can assess how this alters the evolutionary track a galaxy is on, and how the environment links into it.

\section{Results}
\label{sec:Results}

The following section will provide the results of this study, beginning with a short reference to the work done in \cite{Rowntree2026} and the computation of the skeletons that we use, and how they relate to $Z_g$ and the MZR, for the full explanation and assessment of these initial results refer to the previous paper.

Using the aforementioned T-ReX filament finding algorithm and our specified methodology, we compute a skeleton that traces the underlying cosmic structure present in the galaxy distribution. We compute one of these skeletons at 9 snapshots between z=0.625 and 5. 

\begin{figure}
    \centering
    \includegraphics[scale=0.45]{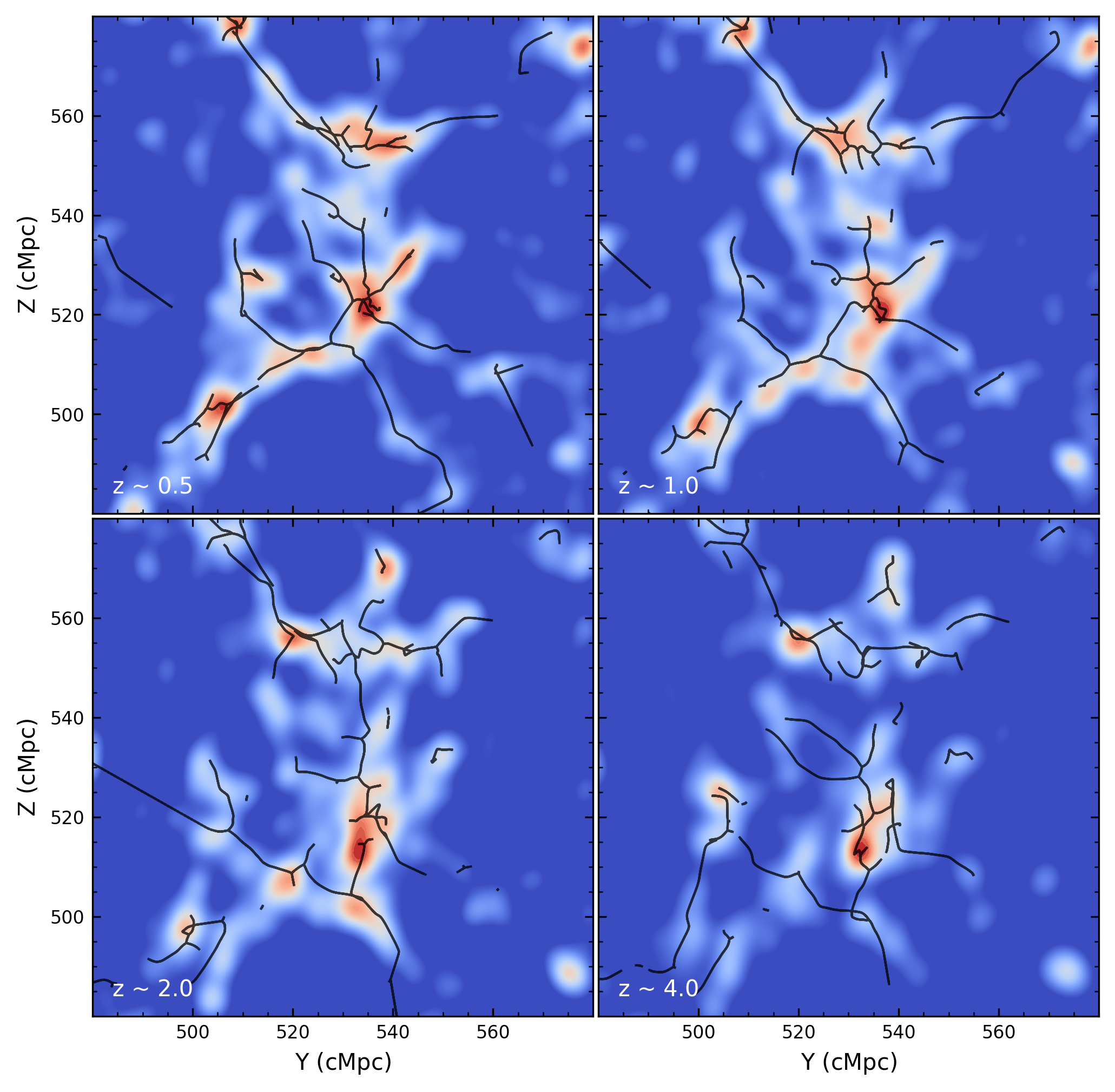}
    \caption{The skeletons produced by {\tt T-ReX} at 4 separate snapshots. The blue to red shaded regions demonstrate the computed local galaxy density at each point, blue being less dense, and red being more dense. The black lines overlaid on this demonstrate the skeleton from the computation. Qualitatively, we can see that the skeleton structure produced by {\tt T-ReX} reasonably traces the galaxy density distribution. This figure is taken from \protect\cite{Rowntree2026}.}
    \label{fig:SkeletonComparison}
\end{figure}

Fig. \ref{fig:SkeletonComparison}, taken from \cite{Rowntree2026}, shows these skeletons at 4 different snapshots. On a purely qualitative basis, we can see that the algorithm is detecting the over-dense regions and building the skeleton to pass through them. Higher redshifts do show a slightly less clean skeleton, however it is still tracing the relevant regions that exist, even in a less continuous density field. The skeletons at each redshift allow us to populate the three environments, and begin to look at the impact they have on the MZR. 

\begin{figure}
    \centering
    \includegraphics[scale=0.3]{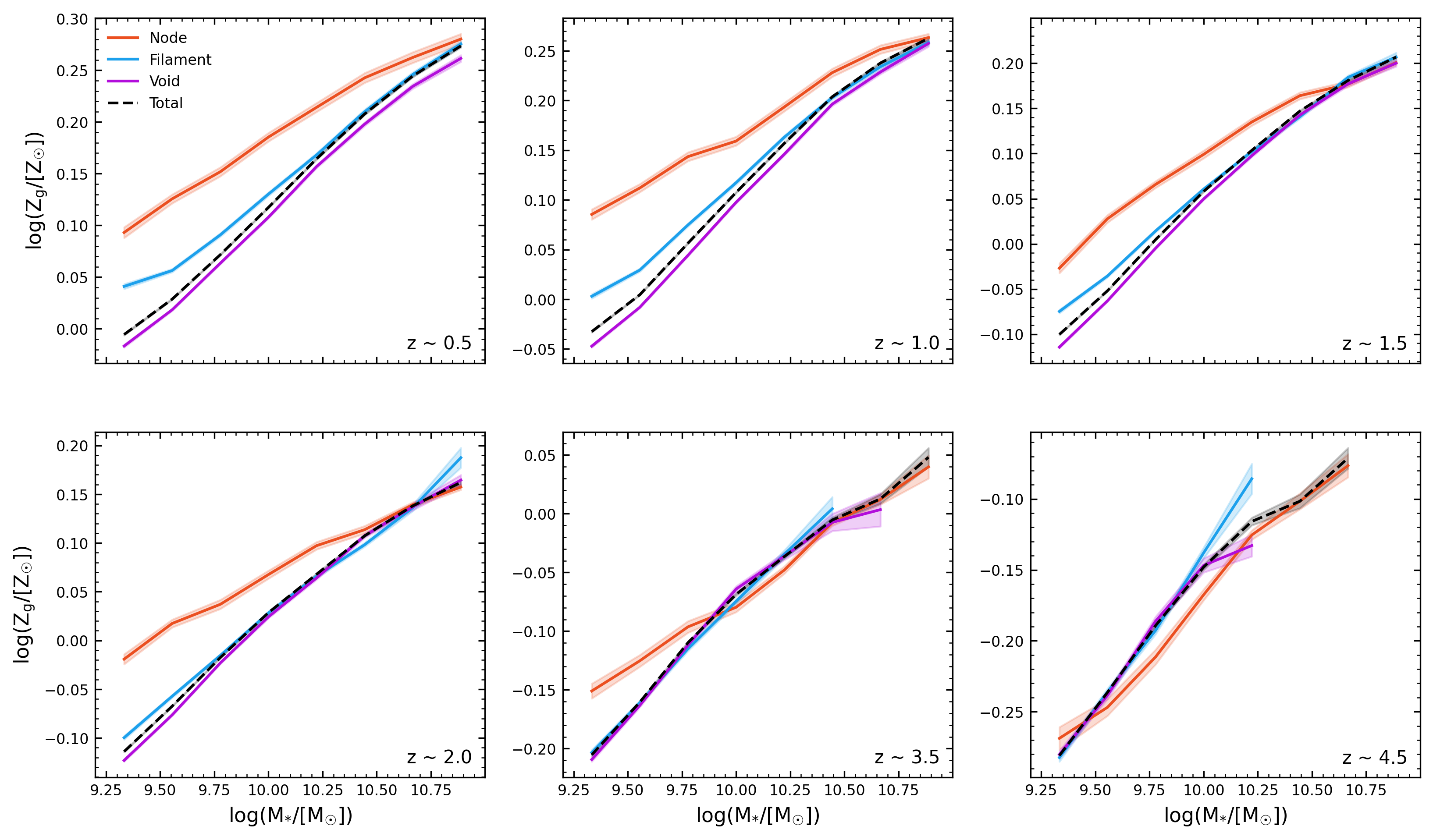}
    \caption{The MZR for node galaxies (orange), filament galaxies (blue), void galaxies (purple), and the total population (black dashed), over 10 consecutive, equal size bins in $\mathrm{M_{\star}}$, for six of the chosen redshift snapshots. Bins with less than 15 galaxies have been removed. The shaded regions represent the standard error on the median value in each bin. This figure is taken from \protect\cite{Rowntree2026}.}
    \label{fig:MZR_Environments_Evolution}
\end{figure}

Fig. \ref{fig:MZR_Environments_Evolution}, taken from \cite{Rowntree2026}, shows the MZR for the total population of galaxies and the 3 individual environments at 6 of the 9 snapshots. At low redshift, the trend due to environment is at its strongest, with nodes showing the highest positive enrichment, particularly for galaxies at low $M_\star$. At high redshift, no trend exists, until $z\approx 3.5$, where the low $M_\star$ galaxies in nodes begin to show the first enrichment. Note that we compare these individual MZRs for each environment to a baseline total MZR for all galaxies to probe the influence each environment has on the MZR itself. It is well known that the MZR itself also has a significant scatter in metallicity, therefore measuring these environments relative to this overall MZR gives an idea of the scatter produced from each environment.
We fit a linear regression to the MZR for the total population of galaxies at each snapshot. By then taking the galaxies in the 3 populations, the nodes, filaments and voids, at each snapshot, we can calculate each galaxy's offset from this linear regression at its particular $M_\star$ and redshift, giving us a value of the MZR residual, $\mathrm{d} \log[Z_g/Z_{\odot}]$, for each galaxy. It is computed as follows in Eq. \ref{eq:residual}.

\begin{equation} 
\label{eq:residual}
\mathrm{d}\log(Z_{g}/Z_{\odot}) = \log\big( \, Z_{g}/Z_{\odot}\big) - \log\langle Z_{MZR}/Z_{\odot} \rangle \,
\end{equation}

This value represents how over or under-enriched a galaxy is relative to the total population of galaxies at its particular $M_\star$ and redshift. Note that due to this, all values of MZR residual given in this work lie on the linear region of the MZR, $9.3 <\approx \mathrm{log[M_\star/M_\odot]} <\approx 10.8$.

\subsection{MZR Residual}

\begin{figure*}
    \centering
    \includegraphics[scale=0.65]{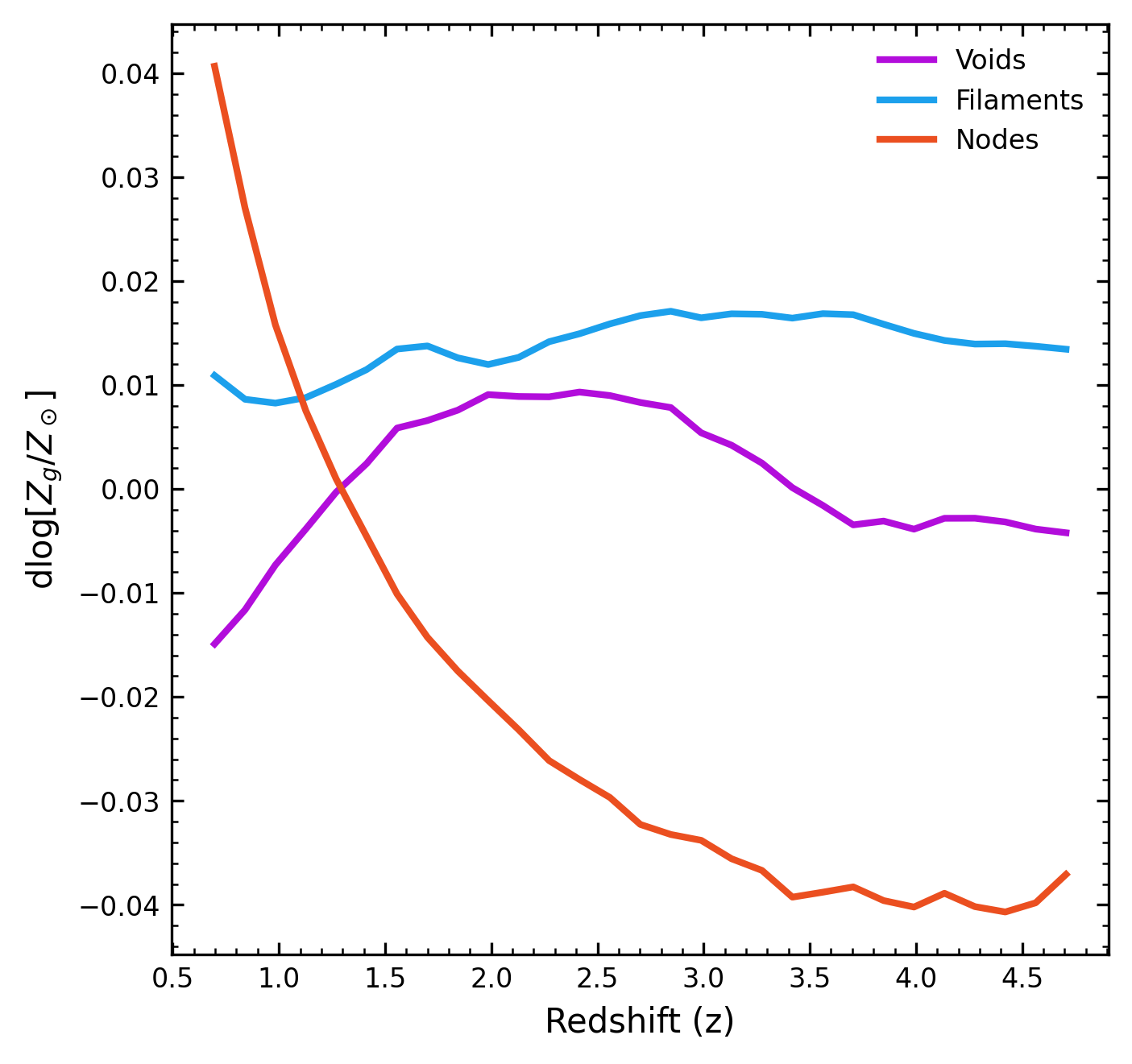}
    \caption{The median MZR residual, $\mathrm{d} \log[Z_g/Z_{\odot}]$, against redshift for the 3 environments. Nodes in orange, filaments in blue and voids in purple. $\mathrm{d} \log[Z_g/Z_{\odot}]$ increases rapidly, from negative to positive, for galaxies that end up in nodes at $z=0.625.$, over the same range, galaxies that end up in voids see a drop in $\mathrm{d} \log[Z_g/Z_{\odot}]$. Note all residual values sit on the linear region of the MZR, $9 <\approx \mathrm{log[M_\star/M_\odot]} <\approx 10.8$}
    \label{fig:MZR_Residual_Redshift}
\end{figure*}

Fig \ref{fig:MZR_Residual_Redshift}. shows the median MZR residual at each snapshot in time. At $z=0.625$, our most present snapshot, the point on the line represents the median $\mathrm{d} \log[Z_g/Z_{\odot}]$ of galaxies within each environment. The next point on the line then represents the median $\mathrm{d} \log[Z_g/Z_{\odot}]$ of these same galaxies one snapshot earlier, so on and so forth.
We can see that in this space, following galaxies back that lie in nodes, filaments and voids at $z=0.625$, there are dramatic differences in $\mathrm{d} \log[Z_g/Z_{\odot}]$ evolution between the environments. Galaxies that end up in nodes, which we know are the most enriched, experience a rapid enrichment below $z=2.2$, in which they go from the most under-enriched population to the most enriched by the present. Filaments as the intermediate environment, see a more static evolutionary track, but still see a small upturn in residual for $z\leq1$. Interestingly, the filament population retain the highest residuals of all environments until approximately $z=1$, in which the nodes rapidly shoot past them. The voids show a flat residual evolution from $z=5$ to $3.8$, at which they then begin to increase in residual until $z=2$, where it peaks, following which we see a falling residual value for the population in the same region that the nodes see their increasing residual. The strongest change in residual occurs below $z=1$ for all environments.

\subsection{Gas Availability/Accretion}

\begin{figure*}
    \centering
    \includegraphics[scale=0.65]{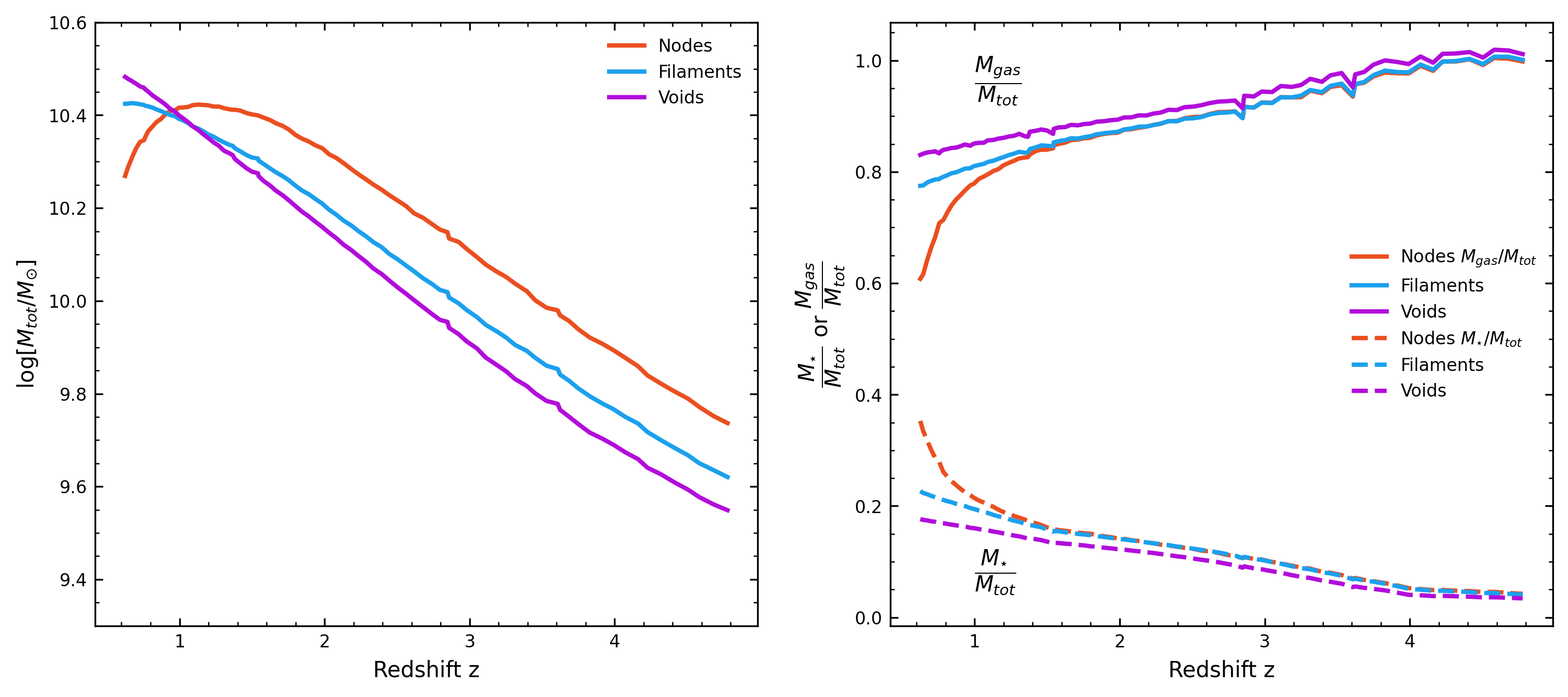}
    \caption{The median evolution of $M_\mathrm{tot}$ (left), $\frac{M_\star}{M_{tot}}$ (right) and gas mass, $\frac{M_{gas}}{M_{tot}}$. (right) across time for each environment. Nodes in orange, filaments in blue, voids in purple. Below $z\approx1$, galaxies that end up in nodes show a turnover in $M_\mathrm{tot}$, with filament galaxies showing signs of following after. This same drop is seen on the right panel as a fall in $\frac{M_{gas}}{M_{tot}}$.}
    \label{fig:Mtot_Mgas_Mstar_Evolution}
\end{figure*}

Fig. \ref{fig:Mtot_Mgas_Mstar_Evolution} shows us the evolution of the total baryonic mass, and mass fractions relative to it, in galaxies across time. In all measures of mass, stellar, gas and total, above $z=1$, galaxies in the nodes display higher median values, filament galaxies display intermediate values, and void galaxies display the lowest. A consistent growth of total mass is seen up to $z=1$, at which the node population begins to decrease in mass, the filaments begin to turnover, and the voids continue to increase in mass. It is below this redshift that the median total baryonic mass is flipped, where void galaxies display the highest, filaments intermediate, and nodes the lowest. By splitting this into fractional $M_\star$ and fractional $M_\mathrm{gas}$, as done in the right panel, we can see that it is infact the change in $M_\mathrm{gas}$ in the system that is contributing to this trend. A $0.21$ dex drop in the gas fraction is observed for node galaxies below $z=1$, concurrently with a $0.15$ dex increase in stellar mass fraction. An offset between these values exists of $0.06$ dex, suggesting that not all of the gas is converted into stars in this time frame, and feedback processes are acting to remove gas from galaxies in nodes below $z=1$. We emphasise that only gravitationally bound, galaxy-associated baryons are counted here.

\begin{figure*}
    \centering
    \includegraphics[scale=0.65]{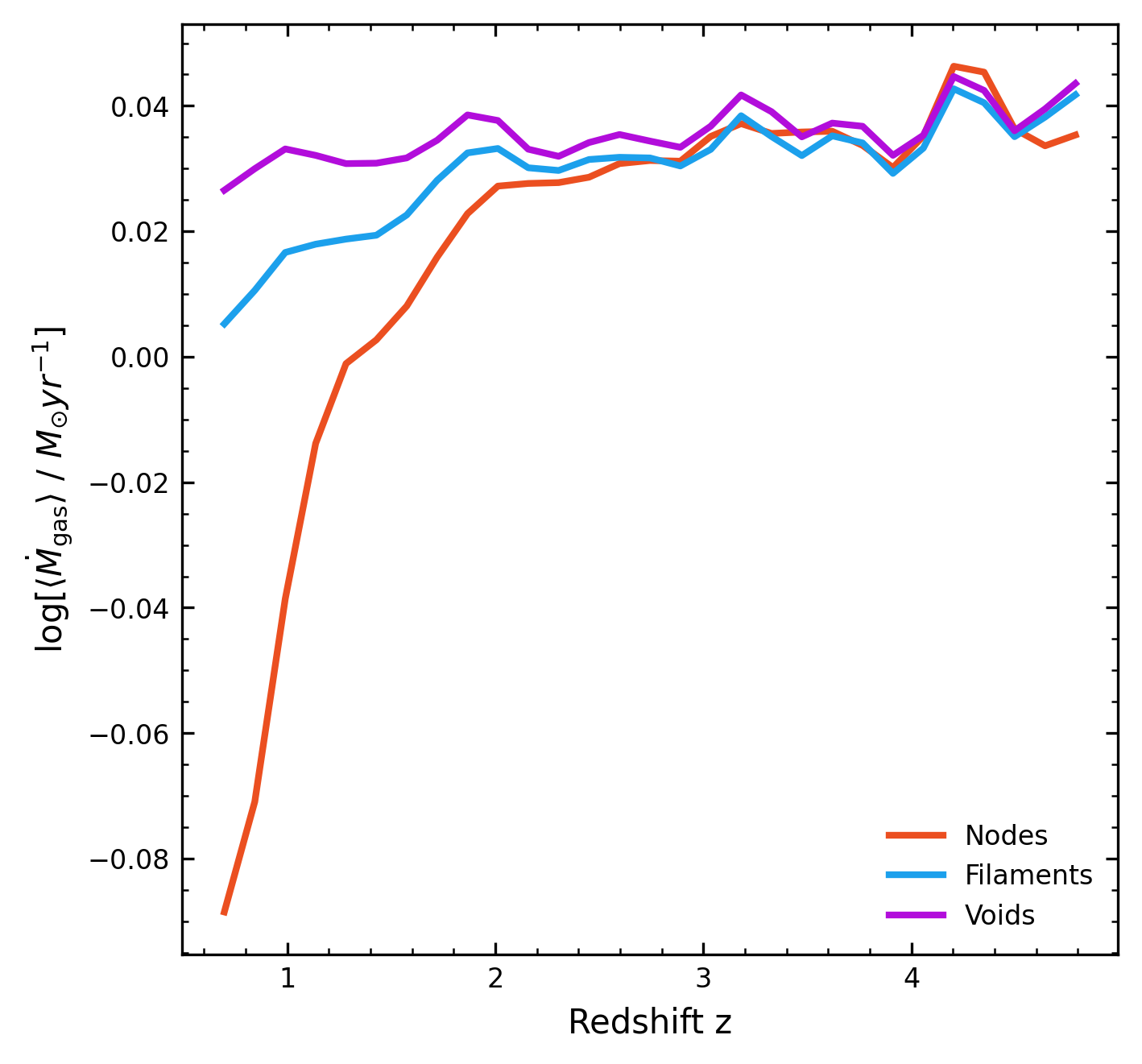}
    \caption{The $\langle \dot{M}_\mathrm{gas} \rangle$ of galaxies in the three environments. Nodes in orange, filaments in blue and voids in purple. Galaxies that end up in nodes show a rapidly falling $M_\mathrm{gas}$ below $z\approx 2$, whilst filament galaxies show a lesser drop, and void galaxies show a flat trend over the same range.}
    \label{fig:dMgasdt_Evolution}
\end{figure*}

Fig. \ref{fig:dMgasdt_Evolution} more closely shows how the gas mass of the galaxies is changing with time, by instead looking at $\langle \dot{M}_\mathrm{gas} \rangle$, the time-averaged rate of change of gas mass. It is clear here that all three environments exhibit a movement towards a negative $\langle \dot{M}_\mathrm{gas} \rangle$, a falling gas mass. The galaxies that end up in nodes reach the turnover in gas mass first, with the filaments showing a similar trend just at a later time, with a possible suggestion from the void trend that they will follow. At $z=0.625$, the Nodes exhibit a rapidly falling gas mass relative to the other environments. Prior to $z=2$, all environments are indistinguishable from one another. The dynamics of the gas in these environments is being shown here, but to show how this affects the MZR across time we look to high and low-mass galaxies in the three environments.

\begin{figure*}
    \centering
    \includegraphics[scale=0.65]{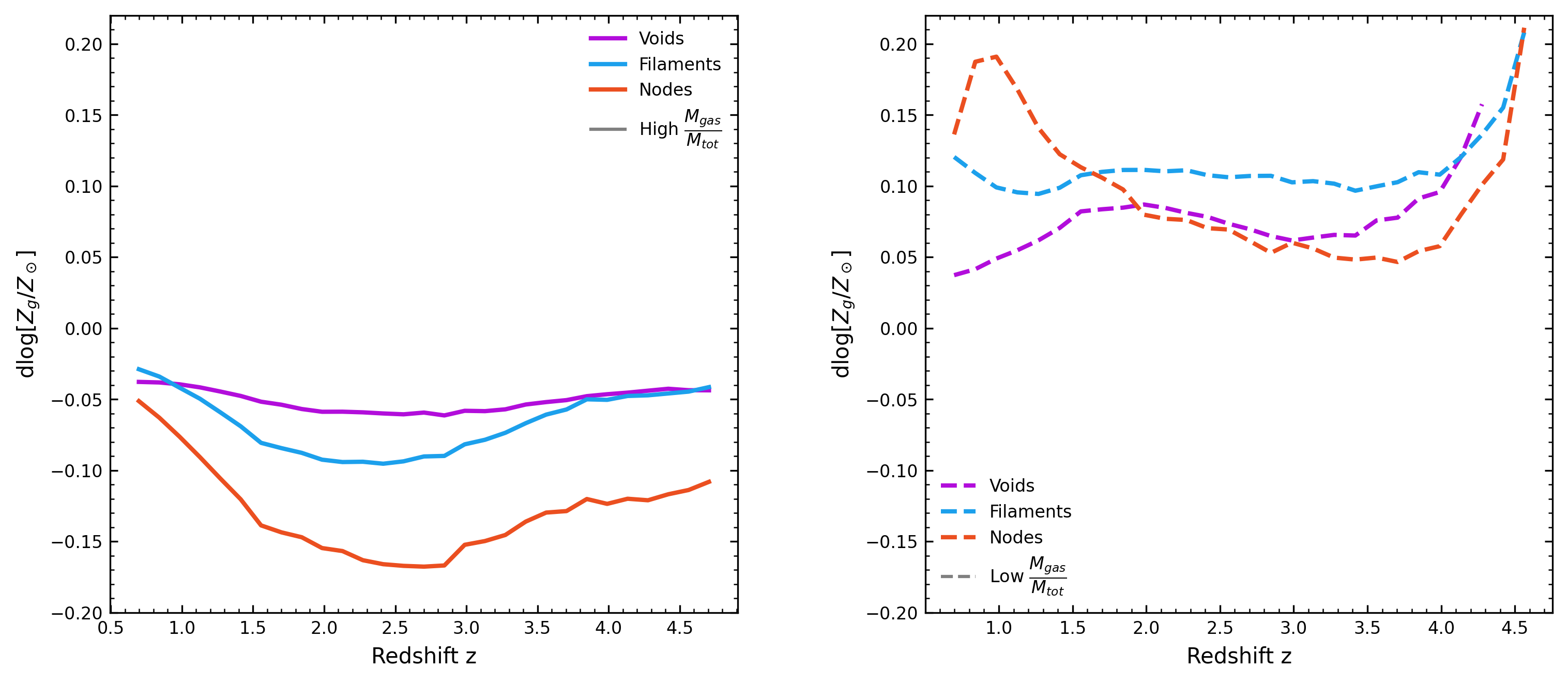}
    \caption{The evolution of $\mathrm{d} \log[Z_g/Z_{\odot}]$ for galaxies in the three environments with high and low fractional gas mass. Node galaxies in orange, filaments in blue and voids in purple. Generally, galaxies with higher gas fraction show reduced $\mathrm{d} \log[Z_g/Z_{\odot}]$, whilst those with low gas fraction show increased $\mathrm{d} \log[Z_g/Z_{\odot}]$, particularly for galaxies that end up in nodes at $z=1$. Note all residual values sit on the linear region of the MZR, $9 <\approx \mathrm{log[M_\star/M_\odot]} <\approx 10.8$}
    \label{fig:Residual_Environment_Mgas_Evolution}
\end{figure*}

Fig. \ref{fig:Residual_Environment_Mgas_Evolution} shows how the evolution of the MZR residual is influenced by the fractional $M_\mathrm{gas}$ of galaxies in the three environments. Galaxies with high fractional $M_\mathrm{gas}$ have negative residuals across all redshifts, an expected result following the process that the accretion of gas regulates the metallicity of galaxies. Galaxies with low gas mass instead show positive values of $\mathrm{d} \log[Z_g/Z_{\odot}]$ for all redshifts. Below $z=1.5$ all environments show an increase in $\mathrm{d} \log[Z_g/Z_{\odot}]$, with the nodes showing the strongest increase, filaments showing a similar increase, then the voids showing a marginal increase. 
Interestingly, looking to galaxies with low fractional gas mass, for $z\leq1.5$, within the same time-frame as the dramatic change in gas mass seen in Fig. \ref{fig:dMgasdt_Evolution}, we see a rapid increase in residual for node galaxies, peaking at approximately $z=1$, then falling afterwards. This is likely due to the late assembly of galaxy clusters.

\begin{figure*}
    \centering
    \includegraphics[scale=0.65]{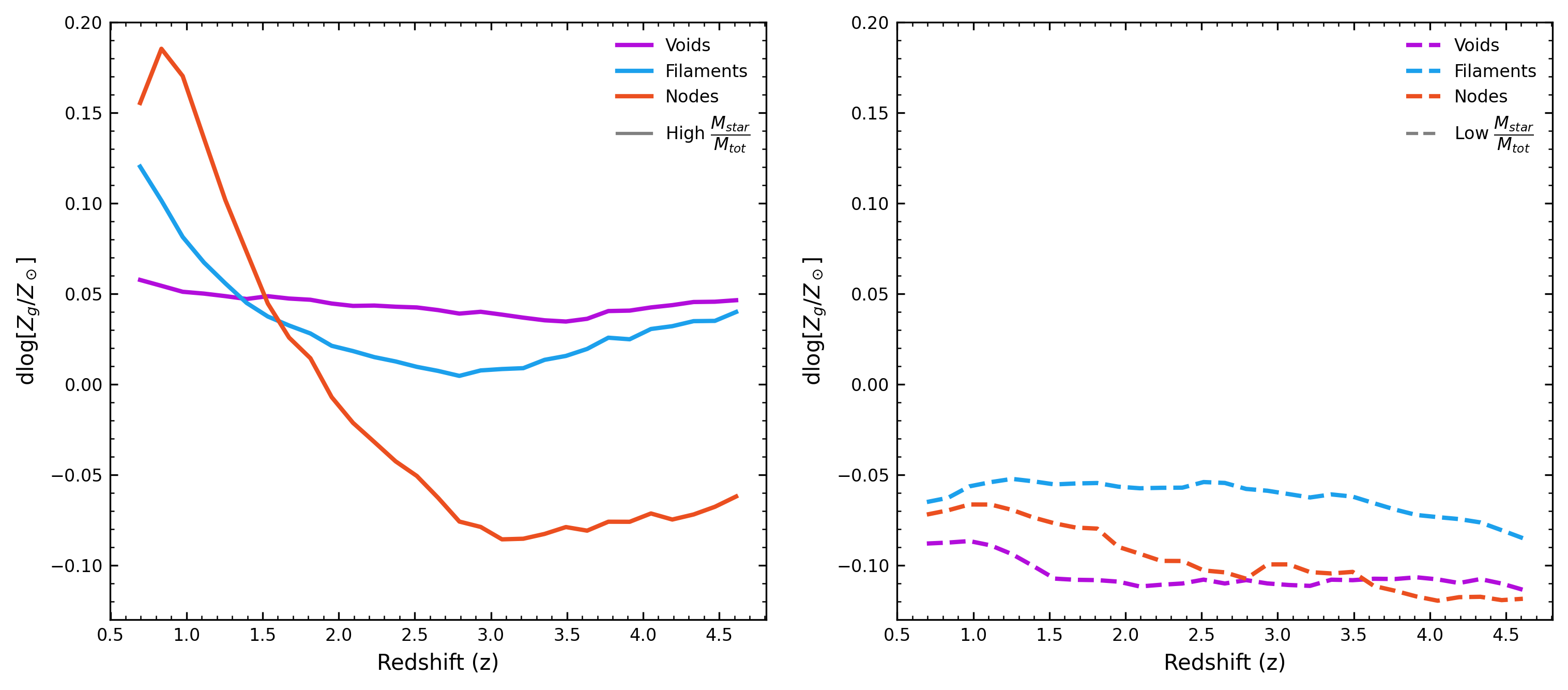}
    \caption{The evolution of $\mathrm{d} \log[Z_g/Z_{\odot}]$ for galaxies with high and low fractional $M_{star}$ in the three environments. Nodes in orange, filaments in blue, and voids in purple. Galaxies with high $M_\mathrm{\star}/M_\mathrm{tot}$ show dramatically increased $\mathrm{d} \log[Z_g/Z_{\odot}]$ values, particularly below $z=1.5$ for galaxies that end up in nodes. This node population also has the largest range of $\mathrm{d} \log[Z_g/Z_{\odot}]$ values across all redshifts, increasing to the lower redshifts, whilst the filament and void population are more flat. Note all residual values sit on the linear region of the MZR, $9 <\approx \mathrm{log[M_\star/M_\odot]} <\approx 10.8$}
    \label{fig:Residual_Environment_Mstar_Evolution}
\end{figure*}

On the other side, Fig. \ref{fig:Residual_Environment_Mstar_Evolution} shows how galaxies with high and low fractional $M_\mathrm{\star}$ evolve in the three environments. Expectedly, galaxies with high fractional stellar mass are the ones that show similar values of residual at $z=0.625$, and show a similar peak to galaxies that have low fractional gas mass, as gas mass typically falls at the expense of increasing stellar mass. The significant difference to the evolution in Fig. \ref{fig:Residual_Environment_Mgas_Evolution} is that galaxies with high fractional $M_\mathrm{gas}$ that end up in nodes, on average, have negative residuals across the full redshift range. The galaxies at low redshift are likely acquiring gas mass from the environment, and end up with high $M_\star$ by the final snapshots. At $z=3$, galaxies with high fractional $M_\mathrm{\star}$] that end up in nodes see a rapid increase of residual up until $z=0.625$. At early times, these galaxies show the lowest residuals between all environments before $z=1.5$, as also seen in Fig. \ref{fig:MZR_Residual_Redshift} where the curve in $\delta \mathrm{log}[Z_{g}]$ for nodes is below filaments and voids until $z=1.5$. Galaxies with low fractional $M_\mathrm{\star}$ show only negative residuals across all redshifts, with filaments and voids showing a relatively flat trend, whilst the nodes show a slight increase of approximately $0.1$ dex over the whole redshift range. These static properties of a galaxy do provide interesting insights, but it is also important to consider how the dynamics of these properties affect a galaxy's enrichment. 

\begin{figure*}
    \centering
    \includegraphics[scale=0.65]{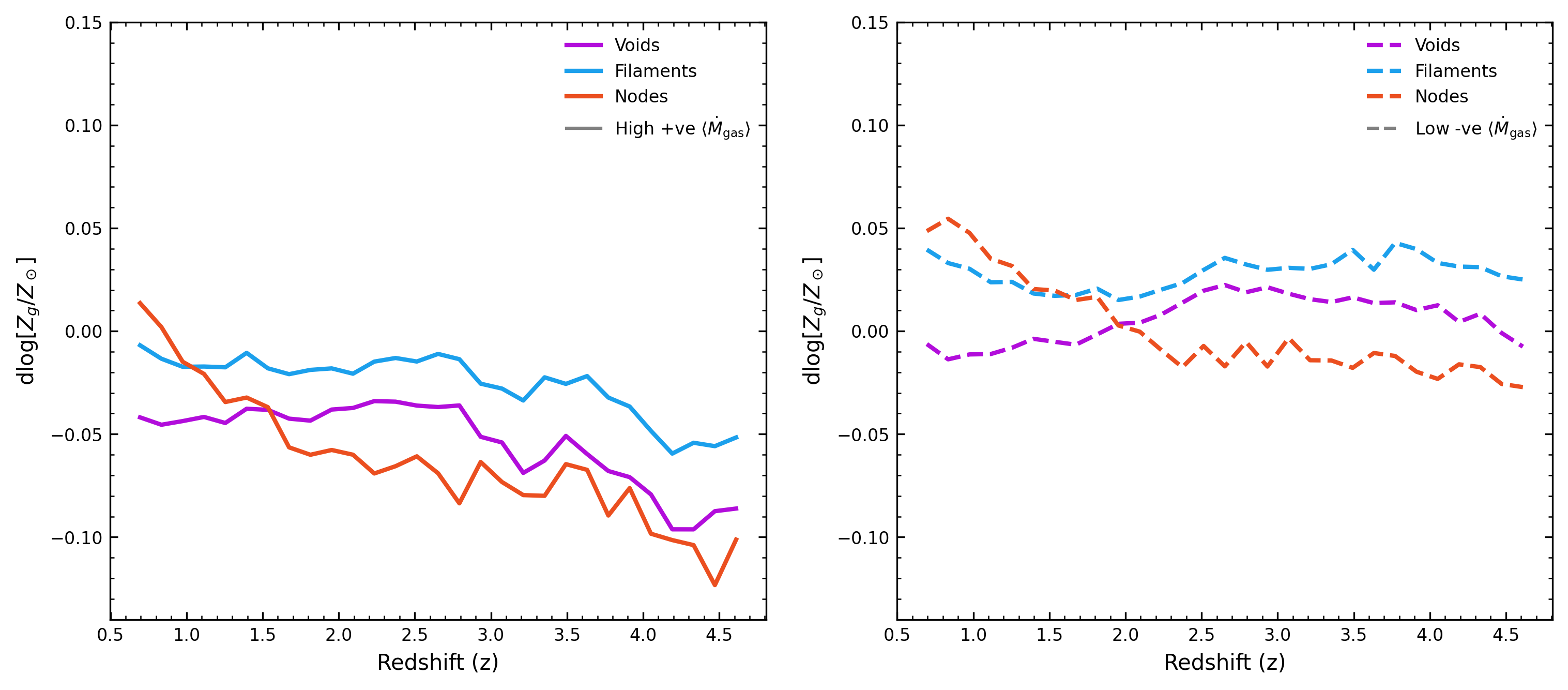}
    \caption{The evolution of $\mathrm{d} \log[Z_g/Z_{\odot}]$ for galaxies in the three environments that are rapidly increasing in gas fraction, left, and rapidly decreasing in gas fraction, right. The nodes are shown in orange, the filaments are shown in blue, and the voids are shown in purple. Galaxies that are losing gas generally show higher, positive $\mathrm{d} \log[Z_g/Z_{\odot}]$, whilst galaxies that are gaining gas show reduced, negative $\mathrm{d} \log[Z_g/Z_{\odot}]$. Note all residual values sit on the linear region of the MZR, $9 <\approx \mathrm{log[M_\star/M_\odot]} <\approx 10.8$}
    \label{fig:Residual_Environment_dMgasdt_Evolution}
\end{figure*}

By looking at Fig. \ref{fig:Residual_Environment_dMgasdt_Evolution} we can see how the MZR residual of the three environments is impacted by creating two populations of galaxies with high and low time-averaged change in fractional gas mass, $\langle \dot{M}_\mathrm{gas} \rangle$. If this value is positive, then a galaxy is gaining gas relative to its total mass, and vice versa for a negative value. Galaxies losing gas relative to their total mass see systematically higher residuals across the full redshift range. These galaxies show very little evolution with redshift, apart from galaxies in nodes below $z=2.0$, where they show a $0.07$ dex increase in $\mathrm{d} \log[Z_g/Z_{\odot}]$ below this redshift.

Within HR5, particularly for large halos containing multiple galaxies, satellite galaxies may lose a significant fraction of their gas soon after infall. In this case, a separation between satellites and centrals is created in this space. An important next step for studies that come after this would be to split the population into the two galaxy types, particularly in the node environments, to assess their individual contributions to the trends we present.

\subsection{Galaxy Mergers}

To begin our analysis into the physical processes, we split our galaxy mergers into two populations, major and minor. We take this cut based on the mass ratio of the two most massive galaxies involved in the merger. A major merger is one where this mass ratio is larger than $0.3$, and a minor merger is one where the mass ratio is lower than $0.3$. As noted, $M_{tot}$ in this work encompasses the total baryonic matter in a subhalo, not the DM. At each snapshot we create three populations: galaxies undergoing major mergers, minor mergers, or no merger at all. We can follow these three populations back across redshift, in the three environments, to see the impact of galaxy mergers on the residual from the MZR, and on a galaxy's chemical evolution.

\begin{figure*}
    \centering
    \includegraphics[scale=0.65]{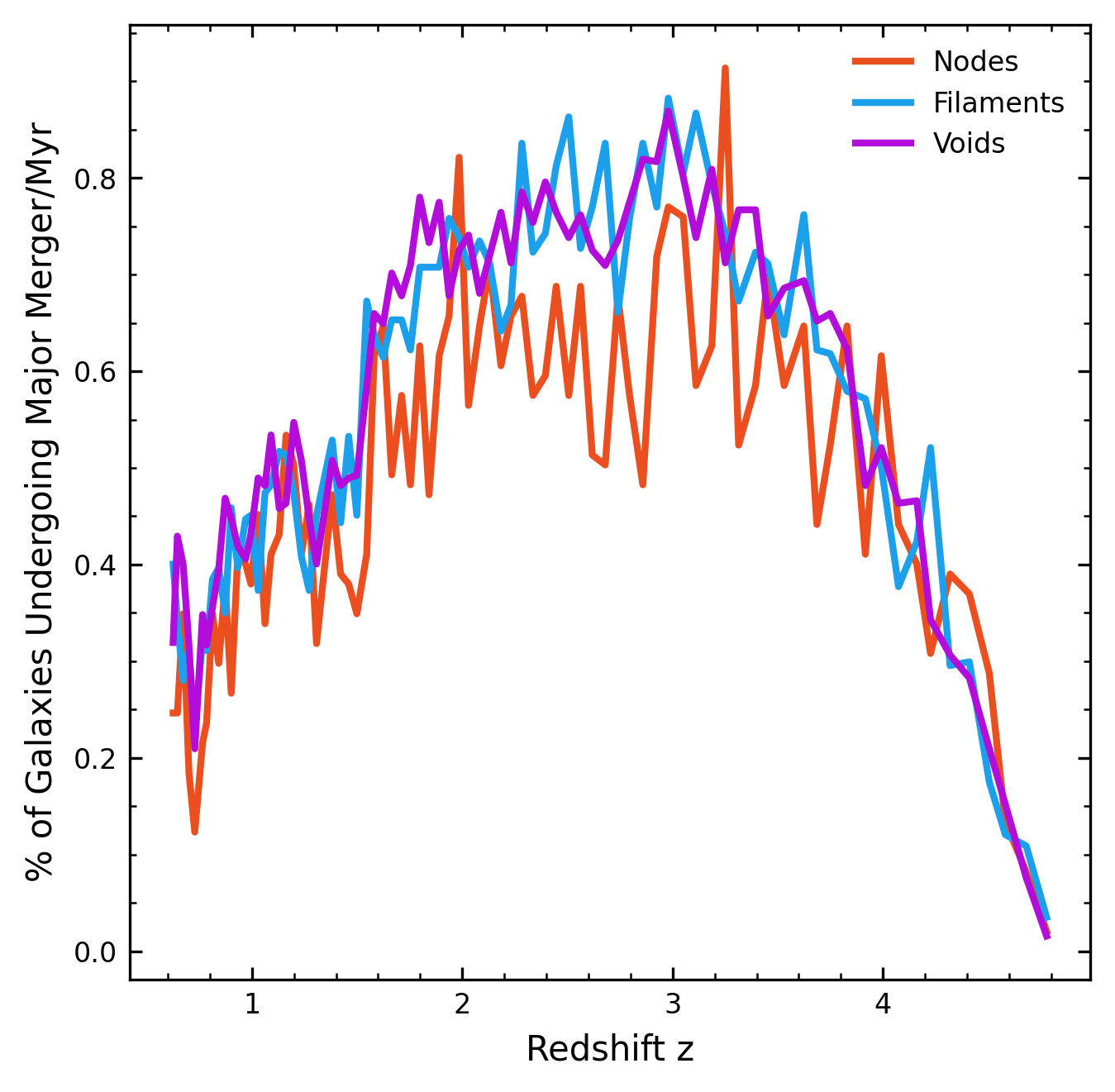}
    \caption{The per cent of galaxies undergoing major mergers at each snapshot in time. Each coloured line represent galaxies that end up in different environments at $z=0.625$. Nodes in orange, filaments in blue, and voids in purple. Irrelevant of environment, all galaxies show similar frequency of major mergers, peaking at approximately $z=2.5$.} 
    \label{fig:Major_Merger_Rate_Evolution}
\end{figure*}

We can first define the percentage of galaxies undergoing mergers at each snapshot in each environmental category, and then plot this percentage across time. Fig. \ref{fig:Major_Merger_Rate_Evolution} shows the peak frequency of major mergers to be occurring at approximately $z=2.7$. There is also no trend observed here between environments, with major mergers occurring at similar frequencies across the definitions. We note that studies like \cite{Fakhouri2009} and \cite{Sureshkumar2024} find that major merger rate increases with local environment, focussing on the analysis of local galaxy density estimates, however here we are analysing the changes in the average merger rate depending on the global environment, focussing on the estimated position of global structures, which is at this moment not well studied.

\begin{figure*}
    \centering
    \includegraphics[scale=0.65]{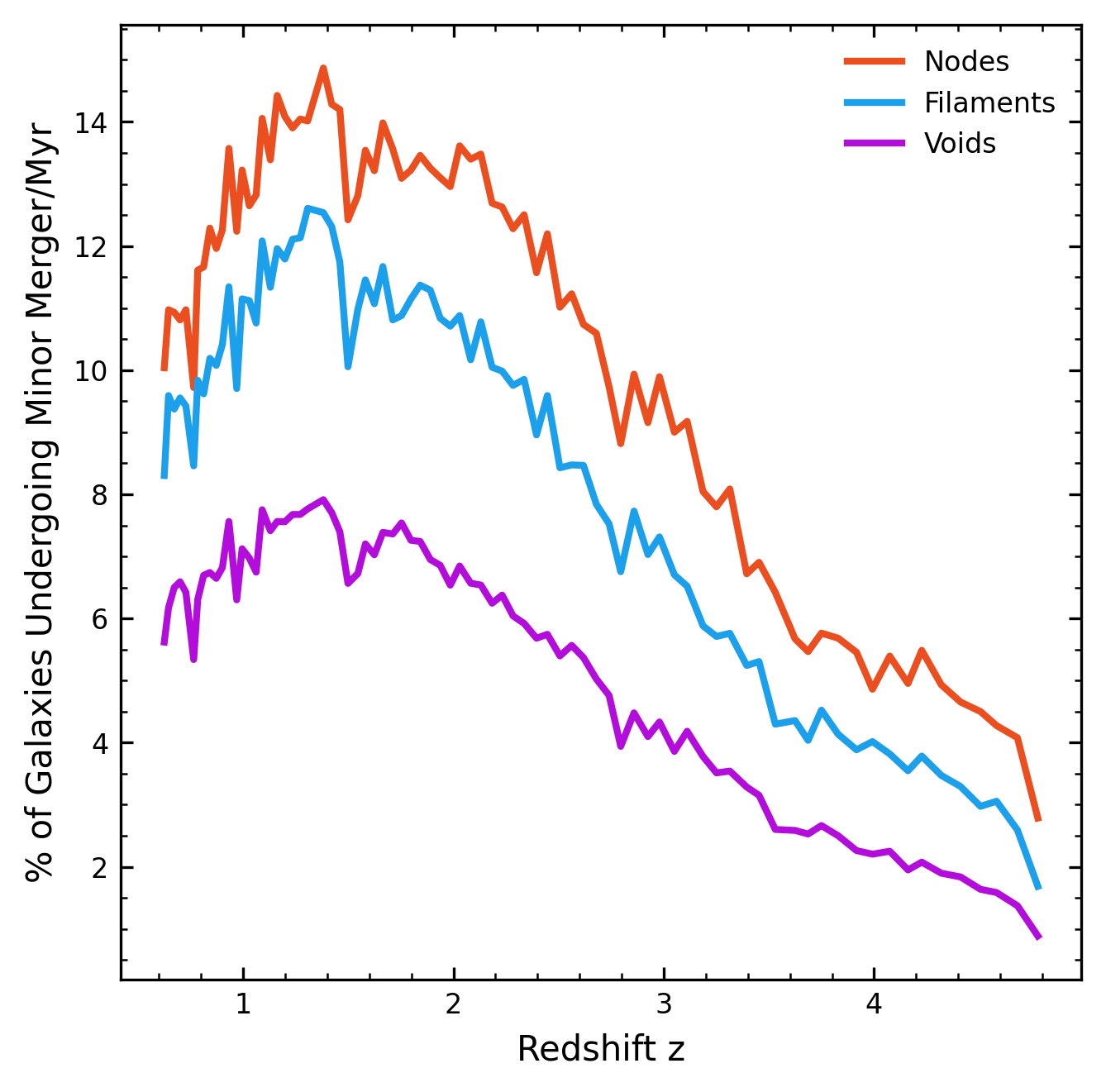}
    \caption{The percent of galaxies undergoing minor merger at each snapshot in time. Each coloured line represent galaxies that end up in different environments at $z=0.625$. Nodes in orange, filaments in blue, and voids in purple. Conversely to major mergers, minor mergers do show a significant environmental dependence in frequency. Galaxies that end up in nodes show a higher frequency of minor mergers, filaments show an intermediate frequency, and voids show the lowest frequency.}
    \label{fig:Minor_Merger_Rate_Evolution}
\end{figure*}

Interestingly, Fig. \ref{fig:Minor_Merger_Rate_Evolution} shows the same evolution however, for minor mergers, where we see that the peak of merger frequency has shifted to later times. The peak of minor mergers instead lies at approximately $z=1.4$, and now is showing a systematic difference in frequency between the environments. All environments show an increasing minor merger frequency between $z=5$ and $z=1.4$, after which the minor merger frequency begins to drop. Galaxies that end up in nodes see higher minor merger frequencies across all of time, peaking at $15\%$ of galaxies undergoing a minor merger at $z=1.4$. The filament population shows intermediate minor merger frequencies across time, peaking at the same redshift but at a lower frequency of $12.5\%$. The void population shows the lowest minor merger rate at all times, peaking at approximately $8\%$ at again the same redshift. Expectedly, minor mergers are clearly dramatically more common than major mergers, and are far more likely in high density regions. However, as the environmental cut is only made at $z=0.625$, this does not mean that these galaxies are in high density clusters at earlier times, yet still the population boasts significantly higher minor merger frequencies, with the trend following the density difference in the environments.

\begin{figure*}
    \centering
    \includegraphics[scale=0.65]{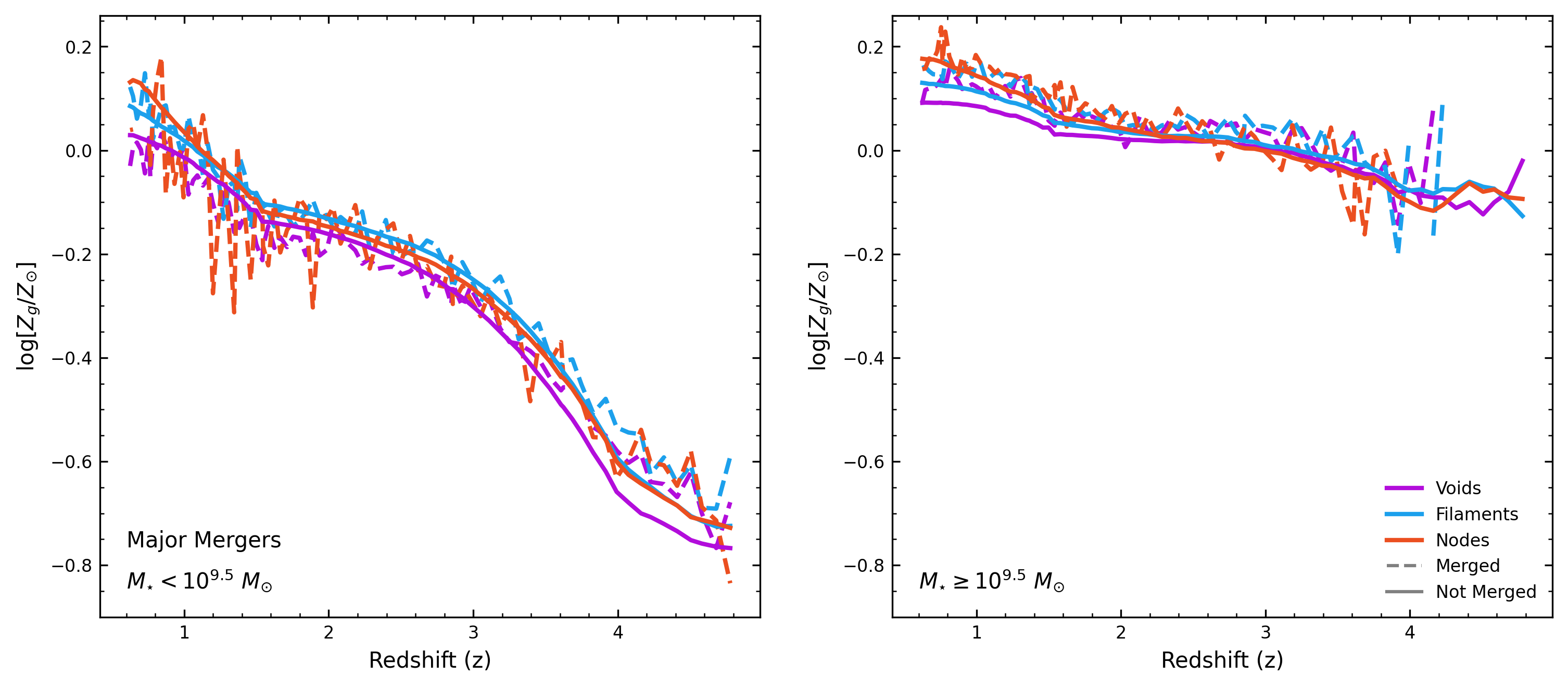}
    \caption{The impact of major mergers on the evolution of $Z_\mathrm{g}$ across time for galaxies that end up in nodes, orange, filaments, blue, and voids, purple. Values of gas metallicity are calculated as the running median across redshift. The solid lines show galaxies that are currently experiencing major mergers, whilst the dashed line shows galaxies that are not currently merging. Although the trends are messy where galaxy population counts run out, major mergers show little impact on $Z_\mathrm{g}$ independent of environment, redshift and $M_\mathrm{\star}$}
    \label{fig:Major_Merger_Rate_Zgas_Evolution}
\end{figure*}

To understand how the two kinds of mergers impact the chemical evolution of galaxies in the different environments, we plot the gas metallicity of galaxies that are undergoing major mergers in the three environments across time, then comparing them to those not undergoing mergers. We also split these populations into $M_\mathrm{\star} \geq 10^{9.5} M_\odot$ and $M_\mathrm{\star} < 10^{9.5} M_\odot$ to investigate a mass dependence of this relationship.
Fig. \ref{fig:Major_Merger_Rate_Zgas_Evolution} shows this for the major merger population relative to non merging galaxies. Note that these trends, particularly for high $M_\mathrm{\star}$, are messy due to the lack of higher mass galaxies in the early universe, even with this, it is clear that for all environments we do not see a significant change in $Z_g$ across all snapshots for galaxies undergoing major mergers.

\begin{figure*}
    \centering
    \includegraphics[scale=0.65]{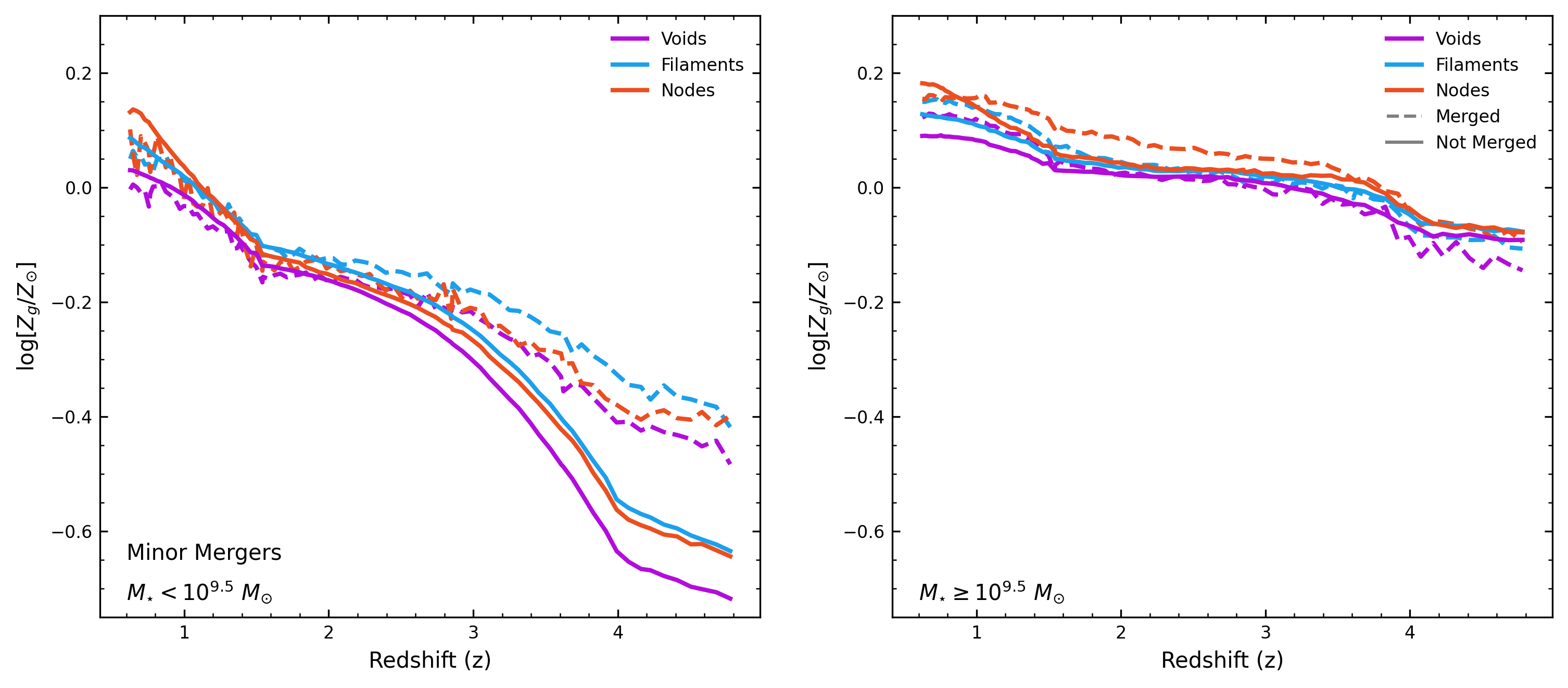}
    \caption{The impact of minor mergers on the evolution of absolute $Z_\mathrm{g}$ across time for galaxies that end up in nodes, orange, filaments, blue, and voids, purple. Values of gas metallicity are calculated as the running median across redshift. The dashed lines show galaxies that are currently experiencing minor mergers, whilst the solid line shows galaxies that are not currently merging. The left panel shows lower mass galaxies with $M_\mathrm{\star} < 10^{9.5} \ \text{M}_{\odot}$, whilst the right panel shows higher mass galaxies with $M_\mathrm{\star} \geq 10^{9.5} \ \text{M}_{\odot}$. Minor mergers show a dramatic impact on the $Z_{g}$ of lower mass galaxies in all environments at early times.}
    \label{fig:Minor_Merger_Rate_Zgas_Evolution}
\end{figure*}

Instead in Fig. \ref{fig:Minor_Merger_Rate_Zgas_Evolution}, a dramatic $0.23$ dex difference is seen at early times for low $M_\mathrm{\star}$ galaxies undergoing minor mergers across all environments. Low $M_\mathrm{\star}$ galaxies that are undergoing minor merger at early times show increased gas metallicities. The magnitude of this offset in metallicity between merging and non-merging galaxies decreases with falling redshift to $0$ dex for all environments by $z=2$. The presence of this enrichment signal in the early universe for low $M_\mathrm{\star}$ galaxies undergoing minor mergers suggests that minor mergers are a dominant growth mechanism at the beginning of galaxies lives, contributing significantly to their early mass assembly and therefore chemical evolution.

\begin{figure*}
    \centering
    \includegraphics[scale=0.65]{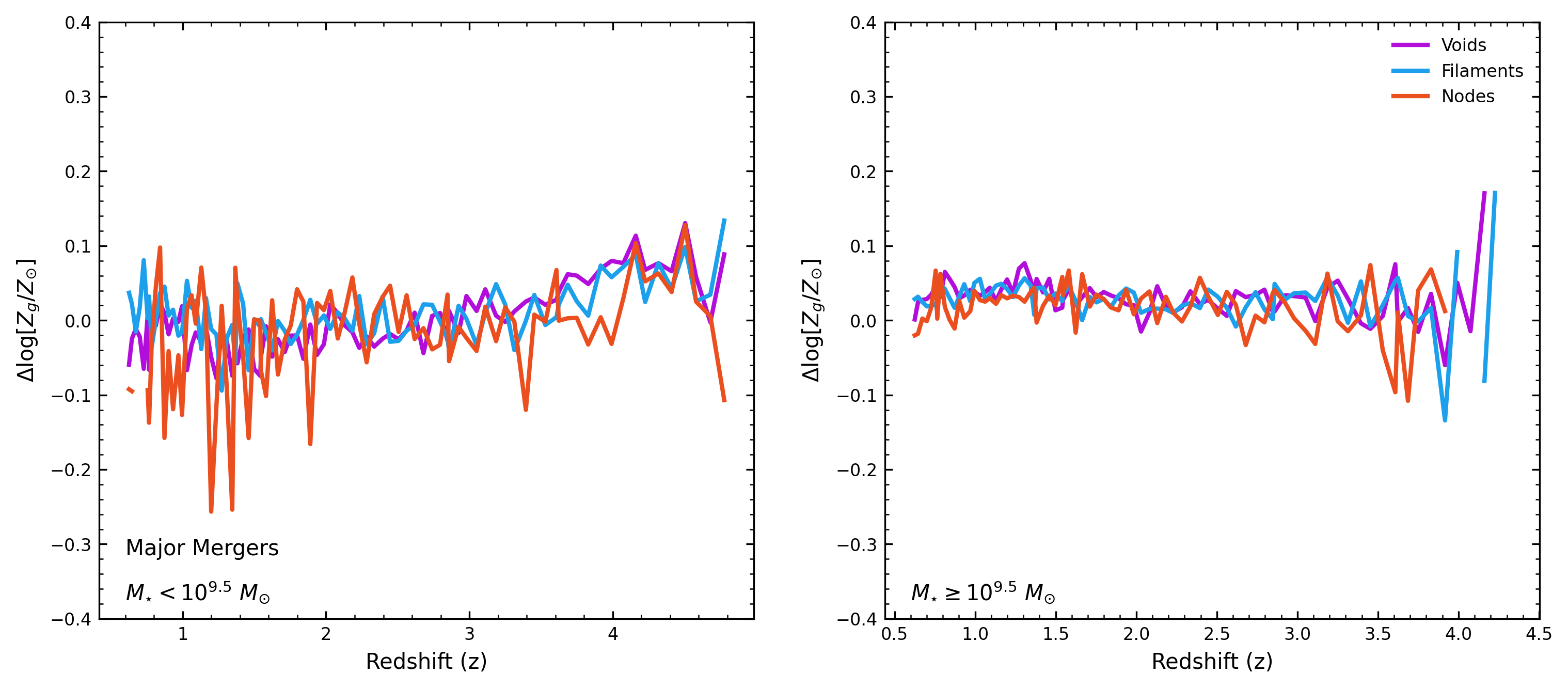}
    \caption{The residual value in absolute $Z_\mathrm{g}$ between major mergers and non-merging galaxies, $\Delta \mathrm{log[Z_{g}/ Z_\odot]}$, across the full range of redshifts in the study. The left panel shows the trend for low $M_\star$ galaxies with $M_\mathrm{\star} < 10^{9.5} \ \text{M}_{\odot}$ whilst the right panel shows the trend for high $M_\star$ galaxies that instead have $M_\mathrm{\star} \geq 10^{9.5} \ \text{M}_{\odot}$. On average, major mergers show no impact on absolute metallicity, with residual values between the populations of near to 0 on average for both the high and low $M_\star$ populations.}
    \label{fig:Major_Merger_Rate_Delta_Zgas_Evolution}
\end{figure*}

\begin{figure*}
    \centering
    \includegraphics[scale=0.65]{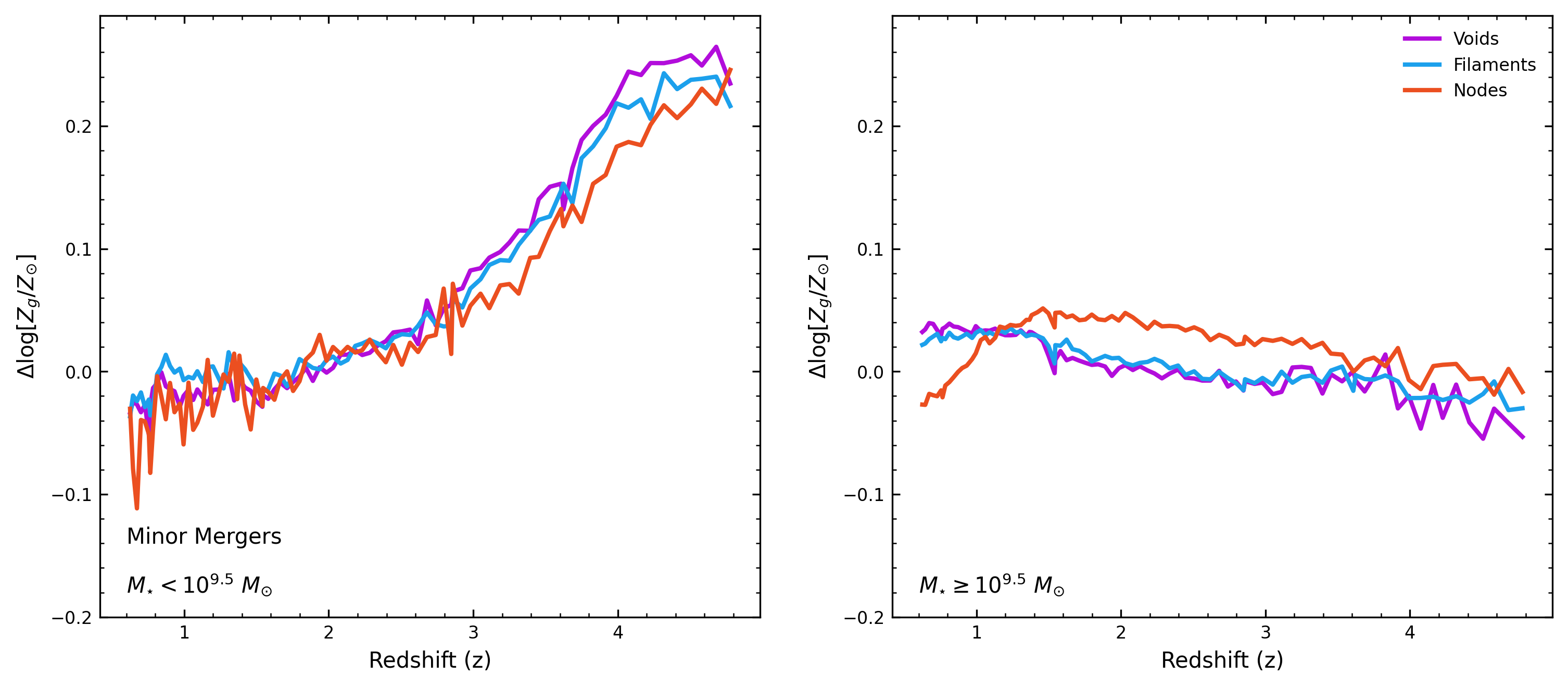}
    \caption{The residual value in absolute $Z_\mathrm{g}$ between minor mergers and non-merging galaxies, $\Delta \mathrm{log[Z_{g}/ Z_\odot]}$, across the full range of redshifts in the study. The left panel shows the trend for low $M_\star$ galaxies with $M_\mathrm{\star} < 10^{9.5} \ \text{M}_{\odot}$ whilst the right panel shows the trend for high $M_\star$ galaxies that instead have $M_\mathrm{\star} \geq 10^{9.5} \ \text{M}_{\odot}$. It is clear that for the low mass galaxy population, minor mergers have a dramatic impact on the absolute metallicity of galaxies in the early universe. The same trend does not exist for galaxies that are not merging.}
    \label{fig:Minor_Merger_Rate_Delta_Zgas_Evolution}
\end{figure*}

To further illustrate this point, we include Fig. \ref{fig:Major_Merger_Rate_Delta_Zgas_Evolution} and Fig. \ref{fig:Minor_Merger_Rate_Delta_Zgas_Evolution} that show the residual absolute $\mathrm{log[Z_{g}/ Z_\odot]}$ between galaxies that are merging and non-merging, $\Delta \mathrm{log[Z_{g}/ Z_\odot]}$. A positive value in this residual indicates that the population of merging galaxies demonstrates higher metallicities, a negative value instead means the merging population shows lower metallicities. It is even more clear in this space that minor mergers in low mass galaxies at early times show increased $Z_g$ relative to galaxies that are not merging.  We also note that there is a small increase in absolute metallicity at intermediate times for higher mass galaxies undergoing minor merger of approximately $0.04$ dex, but this is far smaller than what is observed in the left panel. It is also more obvious in Fig. \ref{fig:Major_Merger_Rate_Delta_Zgas_Evolution} that there is no significant trend for major mergers in either the high or low stellar mass populations, where the values of $\Delta \mathrm{log[Z_{g}/ Z_\odot]}$ are near to 0 on average across the full redshift range.

\begin{figure*}
    \centering
    \includegraphics[scale=0.65]{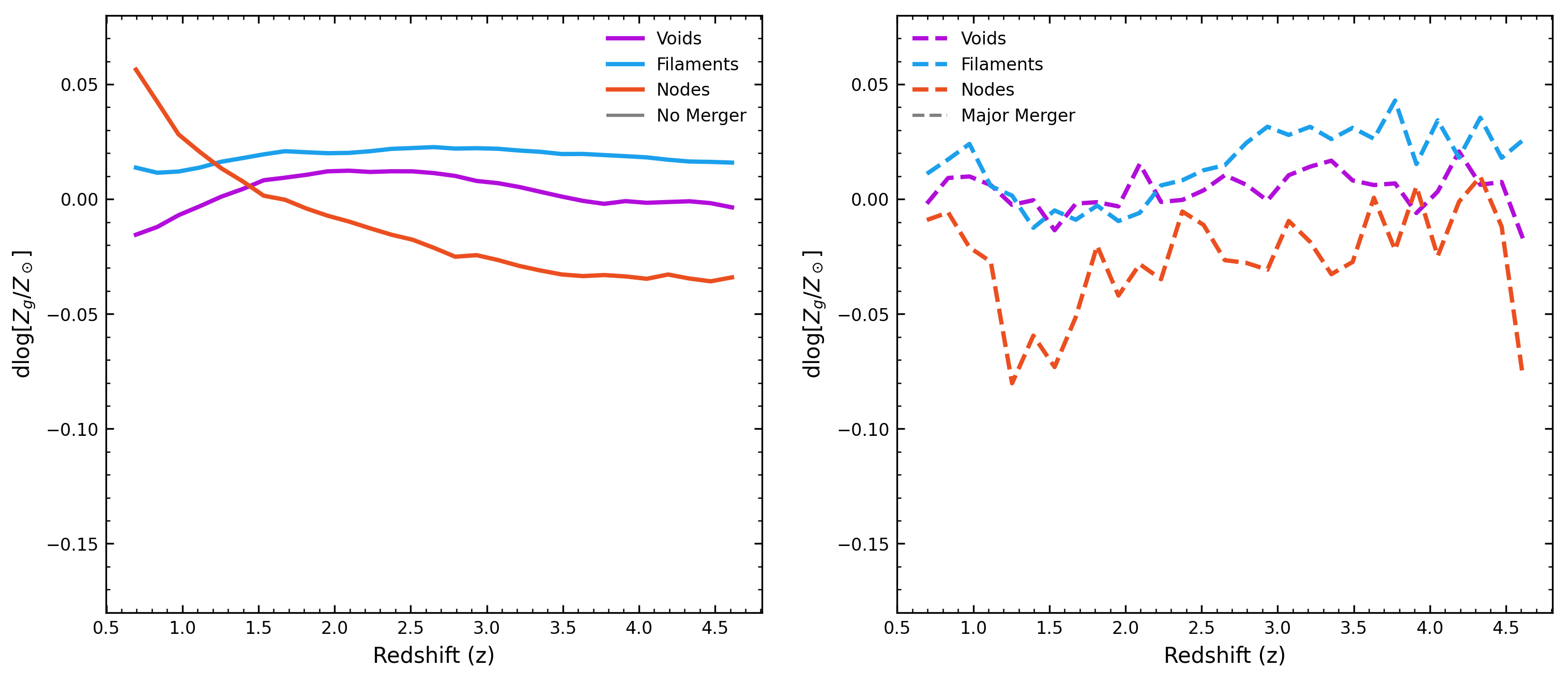}
    \caption{The running median of MZR residual against redshift of galaxies that are not currently undergoing a merger at each snapshot (left) and those that are currently undergoing a major merger (right). Split into populations of 3 environments where galaxies at in the nodes, orange, filaments, blue, and voids, purple at $z=0.625$. The solid lines represent the non-merging populations, whilst the dashed lines represent the population undergoing major merger. Galaxies undergoing major merger show very little change in $\mathrm{d} \log[Z_g/Z_{\odot}]$. Note all residual values sit on the linear region of the MZR, $9 <\approx \mathrm{log[M_\star/M_\odot]} <\approx 10.8$}
    \label{fig:Major_Merger_Rate_Residual_Evolution}
\end{figure*}

It is also important to consider the impact on the scatter in the MZR of minor and major mergers in this context. Fig. \ref{fig:Major_Merger_Rate_Residual_Evolution} shows the MZR residual redshift evolution of galaxies undergoing major mergers, comparing them to those that are not merging. Due to the lower number of majorly merging galaxies at each snapshot, the trends are quite chaotic, however it is again clear that galaxies undergoing major merger do not display significant differences to those that are not in this space. The magnitude of residual for all three environments when undergoing major merger is similar to that of the populations across all snapshots.

\begin{figure*}
    \centering
    \includegraphics[scale=0.65]{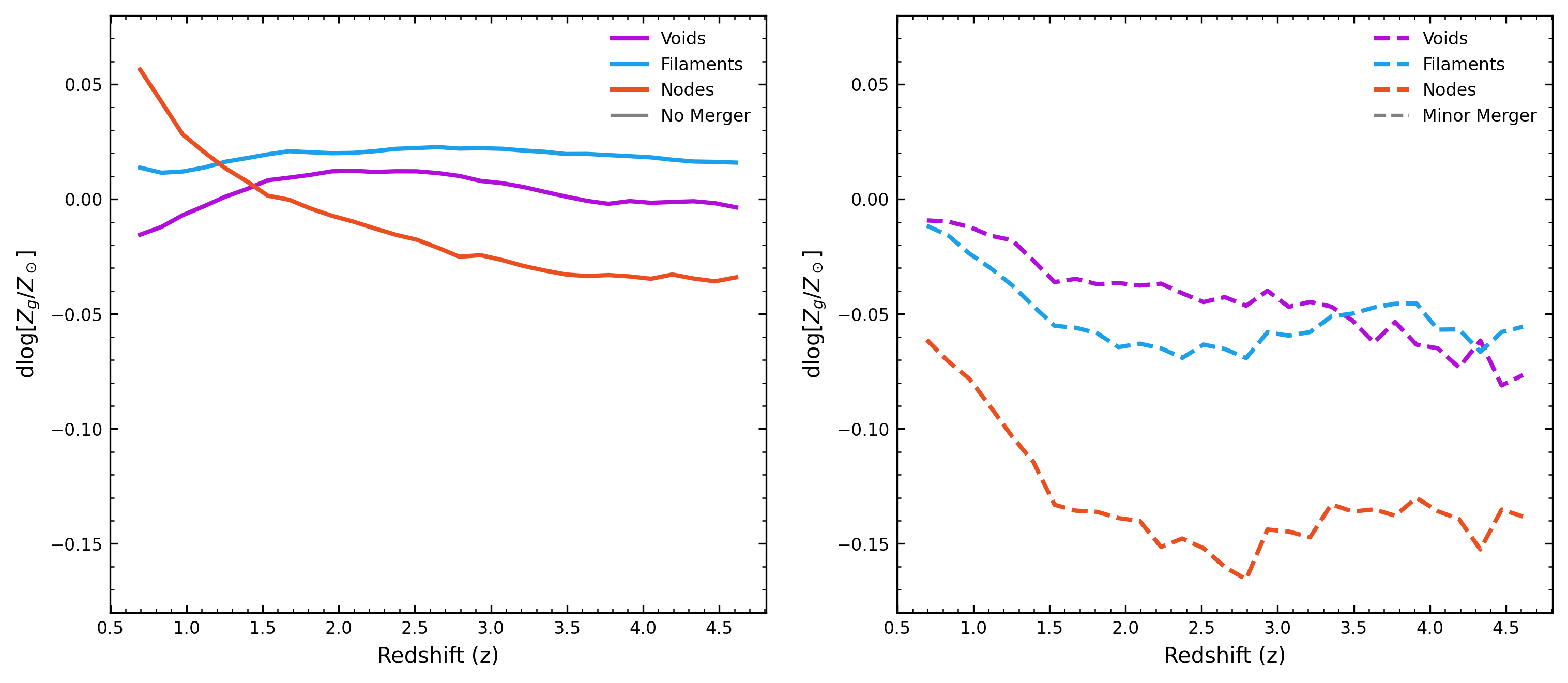}
    \caption{The running median of MZR residual against redshift of galaxies that are not currently undergoing a merger at each snapshot (left) and those that are currently undergoing a minor merger (right). Split into populations of 3 environments where galaxies at in the nodes, orange, filaments, blue, and voids, purple at $z=0.625$. The solid lines represent the non-merging populations, whilst the dashed lines represent the population undergoing minor merger. Galaxies that are undergoing minor merger show significantly reduced and purely negative values of $\mathrm{d} \log[Z_g/Z_{\odot}]$, most strongly for the node population. Note all residual values sit on the linear region of the MZR, $9 <\approx \mathrm{log[M_\star/M_\odot]} <\approx 10.8$}
    \label{fig:Minor_Merger_Rate_Residual_Evolution}
\end{figure*}

Again, in Fig. \ref{fig:Minor_Merger_Rate_Residual_Evolution}, when we consider the minor merger population against $\mathrm{d} \log[Z_g/Z_{\odot}]$, we instead see dramatic difference. Galaxies that are currently experiencing minor merger see reduced MZR residual compared to their non-merging counter parts, and interestingly, they are always negative across all redshifts. The trend between environments is clear, with nodes showing the most reduced residuals as low as $-0.15$, whilst filaments and voids are fairly similar, with the voids being slightly higher in residual, and a reversed trend with environmental density. Interestingly one may interpret this to disagree with the increased gas metallicity for galaxies experiencing minor mergers seen in Fig. \ref{fig:Minor_Merger_Rate_Zgas_Evolution}, however there is no consideration of $M_\star$ here. Fig. \ref{fig:Minor_Merger_Rate_Zgas_Evolution} in the right panel shows the higher mass population of galaxies that have the highest $Z_g$ at this epoch. The average residual calculation of an environment, at a single snapshot, relative to the overall MZR, includes these higher mass galaxies. As such, although these lower mass galaxies undergoing minor merger show significantly increased metallicity, their residual is still negative and reduced relative to the non-merging population that includes these higher mass galaxies. Overall, a scenario is inferred where minor mergers, alongside the gas accretion of unresolved sub-structures, are the dominant mechanisms in the creation of galaxies, driving the evolution of the MZR over time as seen in \citet{Rowntree2026}.

Note that in HR5, there is a galaxy mass resolution limit of $\mathrm{M_{gal}} \approx 2\times10^8 \mathrm{M_\odot}$. This means that it was possible that our minor merger mass ratio of 0.3 was leading to a bias towards higher mass galaxies. To check this we recreated figures that used minor mergers as a dimension, with a minor merger mass ratio of 0.1. All results found in the study either, did not change, or were strengthened by this lower mass ratio, giving confidence in our existing results.

\subsection{SMBH Growth Factor}

The final physical process we investigate is SMBH growth. We do this by looking to the absolute mass of the SMBH, $M_\mathrm{SMBH}$, and its time averaged growth rate, $\langle \dot{M}_\mathrm{SMBH} \rangle$. By looking to high and low values in this parameter and plotting those populations against time, and $\mathrm{d} \log[Z_g/Z_{\odot}]$, gives us insights into how it effects the chemical evolution of galaxies. 

\begin{figure*}
    \centering
    \includegraphics[scale=0.65]{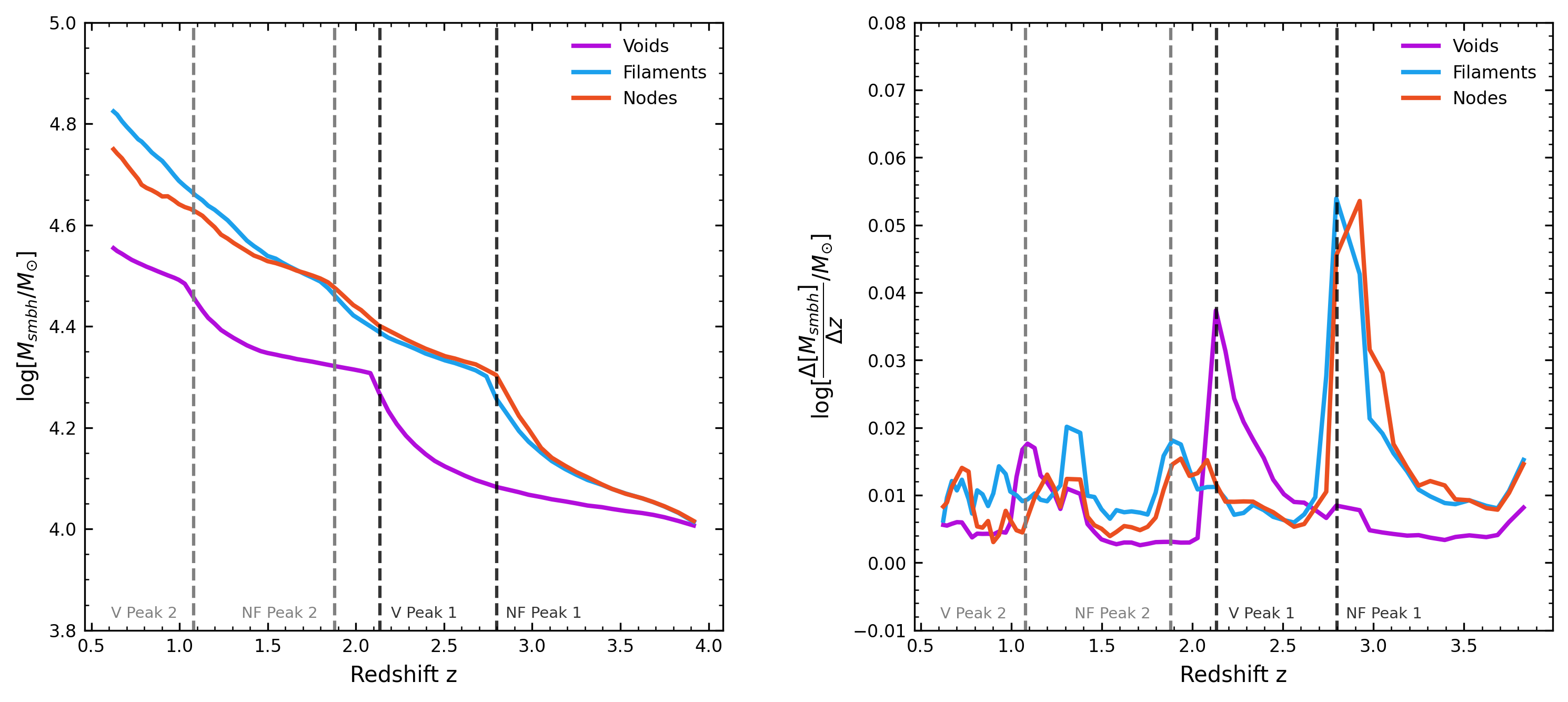}
    \caption{The redshift evolution of the running median $M_\mathrm{SMBH}$ (left) and its differential $\dfrac{\Delta M_{\mathrm{SMBH}}}{\Delta \mathrm{z}}$ (right). Each coloured line represents a unique environment, nodes in orange, filaments in blue, and voids in purple. Galaxies that end up in voids show a significantly different evolution of $M_\mathrm{SMBH}$ than galaxies in the node and filament populations where they increase in mass more slowly. They also show a delayed peak in SMBH growth compared to node and filament galaxies.}
    \label{fig:Msmbh_dMsmbhdt_Evolution}
\end{figure*}

The left panel of Fig. \ref{fig:Msmbh_dMsmbhdt_Evolution} shows the redshift evolution of the $M_\mathrm{SMBH}$ of galaxies in the three environments. The first notable feature of this trend is that the nodes and filament evolutions follow each other closely, whilst the void trend is fully separated at all redshifts. Although it is separated, it still follows the same shape of trend, just offset in time, following a slower evolution. There are two main occurrences where $M_\mathrm{SMBH}$ begins to grow rapidly, until a certain mass is met, at which the growth of the BH is attenuated. The first of these events occurs at $M_\mathrm{SMBH} \approx 10^{4.3} \ \mathrm{M_{\odot}}$, and the second occurs at $M_\mathrm{SMBH} \approx 10^{4.5} \ \mathrm{M_{\odot}}$. Interestingly, the first event occurs at $z=2.8$ for nodes and filaments, whereas it occurs later for voids at $z=2.1$. The second event is also offset for voids, occurring at $z=1.1$, whilst it occurs at an earlier time of $z=1.9$ for the nodes and filaments. Below $z=1.5$, the nodes and filaments begin to deviate from one another, with the filaments actually rising to higher $M_\mathrm{SMBH}$ values than the nodes. Finally, at high redshift, the difference between the environments is at its minimum; this difference increases with time. By looking at Fig.~\ref{fig:Msmbh_dMsmbhdt_Evolution}, in the early stages of BH evolution, the values of $M_\mathrm{SMBH}$ at which the growth rate $\dot{M}_\mathrm{SMBH}$ is attenuated approximately result from BH-BH mergers.
The BH seed mass in HR5 is $10^{4} \ \rm{M_{\odot}}$. The first peak arises at $M_\mathrm{SMBH} \approx 10^{4.3} \ \rm{M_{\odot}}$, corresponding to approximately $2 \times10^{4} \ \rm{M_{\odot}}$, or the mass of a merger of two BH seeds. The second peak arises at $M_\mathrm{SMBH} \approx 10^{4.5} \ \rm{M_{\odot}}$, corresponding to approximately $3 \times10^{4} \ \rm{M_{\odot}}$, possibly a merger of $10^{4} \ \rm{M_{\odot}}$ and $2 \times10^{4} \ \rm{M_{\odot}}$. In HR5, two BHs coalesce when their separation is smaller than $4 \ \mathrm{dx}$, where dx is the resolution limit, and their relative velocity is lower than the escape velocity. However, as mentioned in section 2.4.2, BHs in HR5 can grow through gas accretion. We interpret the gradual rise of $M_\mathrm{SMBH}$ followed by an attenuation of its growth in Fig.~\ref{fig:Msmbh_dMsmbhdt_Evolution} as resulting from the boosted, Eddington-limited Bondi accretion used in HR5 ($\dot{M}_\mathrm{SMBH}\propto M_\mathrm{SMBH}^2$), which self-regulates once feedback evaporates the cold gas that fuels the black hole \citep{Dubois12}.

Another perspective on how $M_\mathrm{SMBH}$ evolves with time is shown in the right panel of the figure. By taking the differential of the left panel, we present the $\Delta M_\mathrm{SMBH}/\Delta t$, or the average SMBH growth factor, of the galaxies in the three environments at each point in time. Here, the peaks are clear in the three environments, appearing as high $\Delta M_\mathrm{SMBH}/\Delta t$ at the specific redshifts. It is more evident here that the environments all follow the same evolution, however it happens slower in the voids. At low redshifts, the growth of the SMBH in nodes and filaments becomes more chaotic. Over all the redshifts, the void population has systematically slower SMBH growth than the other environments.

\begin{figure*}
    \centering
    \includegraphics[scale=0.65]{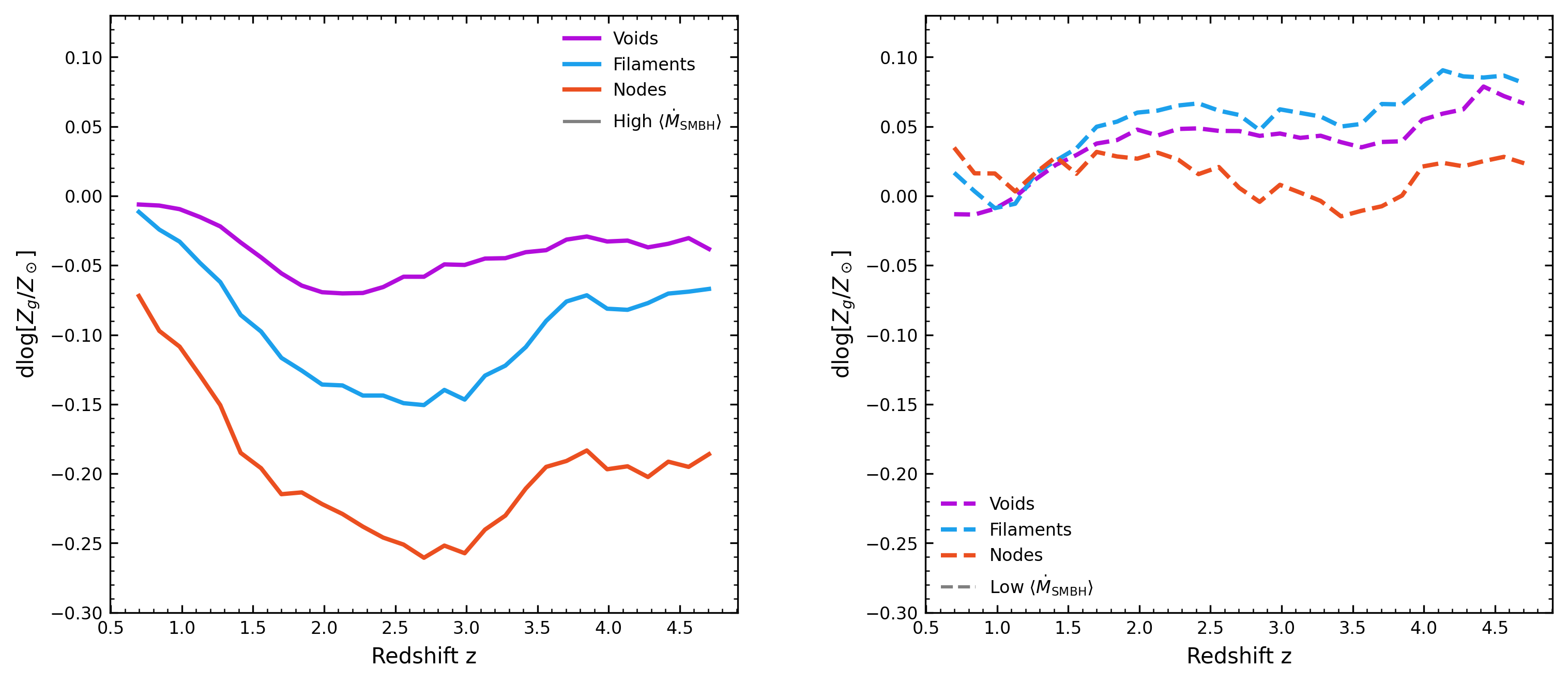}
    \caption{The evolution of the running median of $\mathrm{d} \log[Z_g/Z_{\odot}]$ for galaxies that have high (left, solid line) and low (right, dashed line) $\langle \dot{M}_\mathrm{SMBH} \rangle$. Each individual coloured line represents a different environment, nodes in orange, filaments in blue, voids in purple. Galaxies that have high $\langle \dot{M}_\mathrm{SMBH} \rangle$ show reduced, and only negative, $\mathrm{d} \log[Z_g/Z_{\odot}]$ at intermediate redshifts, between $z=1.5$ and $3.5$. Galaxies in nodes show the offset relative to the overall MZR, followed by the filament population, and then the void population. Below $z=1.5$, the values of $\mathrm{d} \log[Z_g/Z_{\odot}]$ for all environments increase towards 0 again. Note all residual values sit on the linear region of the MZR, $9 <\approx \mathrm{log[M_\star/M_\odot]} <\approx 10.8$}
    \label{fig:Residual_dMsmbhdt_Evolution}
\end{figure*}

Finally we look to the evolution of the MZR residual for galaxies that have high and low SMBH growth between snapshots, $\langle \dot{M}_\mathrm{gas} \rangle$. The left panel of Fig. \ref{fig:Residual_dMsmbhdt_Evolution} shows galaxies in the three environments that display high SMBH growth. Across all redshifts, this population show negative $\mathrm{d} \log[Z_g/Z_{\odot}]$, however it is in the intermediate redshifts, between $z=1$ and $z=3$, where the lowest values of $\mathrm{d} \log[Z_g/Z_{\odot}]$ are seen. It is this period of time where we see the peaks in SMBH growth occur in Fig. \ref{fig:Msmbh_dMsmbhdt_Evolution}. At low redshifts, below $z=1$, $\mathrm{d} \log[Z_g/Z_{\odot}]$ increases again as SMBH growth is seen to wane in the previous figure. In the right panel of this figure, we instead see the galaxies that have slow SMBH growth. Across all redshifts, we see much higher values of $\mathrm{d} \log[Z_g/Z_{\odot}]$, and instead of an increase of $\mathrm{d} \log[Z_g/Z_{\odot}]$ for low redshift, we see a decreasing trend. Interestingly in both panels, the nodes demonstrate the lowest residuals in these spaces. For high $\langle \dot{M}_\mathrm{SMBH} \rangle$, filaments show intermediate values whilst the void galaxies show the highest. For low $\langle \dot{M}_\mathrm{SMBH} \rangle$, the filament and void galaxies are fairly indistinguishable from one another. A final key feature of this plot is the difference between the high $\langle \dot{M}_\mathrm{SMBH} \rangle$ and low $\langle \dot{M}_\mathrm{SMBH} \rangle$ trends in the same environment. For the voids, the least change in $\mathrm{d} \log[Z_g/Z_{\odot}]$ is seen of all environments, with minimal change at $z=0.625$, and the greatest offset relative to the overall MZR of $0.1$ dex in $\mathrm{d} \log[Z_g/Z_{\odot}]$ at intermediate times, $z=2$. For the filaments, there is still minimal change in $\mathrm{d} \log[Z_g/Z_{\odot}]$ at $z=0.625$, however the offset in $\mathrm{d} \log[Z_g/Z_{\odot}]$ relative to the overall MZR at intermediate times has increased to $0.22$ dex, peaking now slightly further back in time at $z=2.5$. The nodes show the largest change in $\mathrm{d} \log[Z_g/Z_{\odot}]$ due to high or low $\langle \dot{M}_\mathrm{SMBH} \rangle$, now actually showing a change at $z=0.625$ of $0.1$ dex, but a large maximum change of $0.26$ dex for intermediate times, yet again the peak offset relative to the overall MZR occurring slightly earlier than the previous environment, at $z=2.8$. We speculate that high SMBH growth here suggests the presence of strong metal-enriched outflows, from the galaxy center, that reduce the metallicity of the overall galaxy which keeps the $\log[Z_g/Z_{\odot}]$ of these galaxies negative, below the MZR. The same pattern in $\log[Z_g/Z_{\odot}]$ could also be expected for a inflow of metal-poor gas to the galactic center where metallicity is regulated all whilst the SMBH undergoes some accretion. Note that in these highly dense regions, it is understood that metal-poor gas is more sparse.

\section{Discussion}
\label{sec:Discussion}

Similar to \cite{Rowntree2026}, this work shows a truly environmentally dependent chemical evolution of galaxies over time. This study demonstrates that the environment, and physical processes operating inside of galaxies, combine to present an overall picture of the chemical evolution of galaxies throughout their lifetime. The physical processes at play in the universe, notably those mentioned in this work, vary significantly between our identified environments and have been shown to have large impacts on the metallicity of galaxies. The results show a scenario where the environment a galaxy resides within decides the type and magnitude of physical processes the galaxy will be exposed to, and from this, its own unique chemical evolution occurs, leaving chemical markers of which processes it has experienced, directly tied to the environment(s) that it has existed within and passed through. We also propose that at different times in the universe's history, different physical processes are acting at different efficiencies and frequencies. 

\subsubsection*{Galaxy Mergers}
When the universe is in its early stages, $z>4$, densities are low, galaxies are low mass, and there is significantly more pristine gas available for star formation. At this time, minor mergers provide the largest difference in absolute metallicity, $\log{Z_{g}/\mathrm{Z_{\odot}}}$, particularly for low $M_\mathrm{\star}$ galaxies, $M_\mathrm{\star} \leq 10^{9.5} \ \mathrm{M_\odot}$. These mergers contribute to the mass assembly of these smaller haloes, that are then involved in the assembly of more massive galaxies. Because more massive galaxies have higher average metallicities at any epoch, we suggest that this is one of the key processes that produces the average MZR that we observe. Simultaneously to this, galaxies that are undergoing minor merger show reduced $\mathrm{d} \log[Z_g/Z_{\odot}]$ for all redshifts compared to those not merging. This agrees with previous ideas proposed in \cite{Rowntree2026} where evidence suggests, through interpretation of $\mathrm{[O/Fe]}$ values, that the accretion of pristine gas is a significant contributor to the regulation of gas metalicity. In this case, a higher stellar mass, higher metallicity galaxy accretes a lower stellar mass, lower metallicity one, which is very similar to the regulatory process of cold-gas accretion. It is these minor mergers that are responsible for the largest increases in absolute metallicity within HR5 below $z=4$, whilst the consistently reduced $\mathrm{d} \log[Z_g/Z_{\odot}]$ over the whole of cosmic time shows how this process is similar to cold-gas accretion, acting to regulate the metallicity of galaxies, all whilst contributing to their mass growth, accelerating their chemical evolution post-event. Note that at this redshift, minor merger frequency is at a minimum. Even though approximately 2\% galaxies are experiencing minor mergers here, the impacts on $\log[Z_g/Z_{\odot}]$ seen due to them are far larger than the other processes at this time. Note that we here are focussed on the magnitude of difference in $\log[Z_g/Z_{\odot}]$ between galaxies undergoing minor merger, and those that are not. We do not propose that minor mergers dominate average trends, just that their impact on the magnitude of $\log[Z_g/Z_{\odot}]$ is larger than the impact due to other processes at these early times.

A deeper understanding of both minor and major mergers and how they impact chemical evolution within the different environments could be acquired by looking to the specific stages of the merging process. Within HR5 we do not have the time resolution to achieve this. Understanding how $Z_g$ changes throughout the full merger process using discrete time steps would allow us to see how global environment itself impacts these specific parts of the total merger process.

Across the full range of redshifts, the reduction in $\mathrm{d} \log[Z_g/Z_{\odot}]$ for node galaxies undergoing minor merger is far larger than the reduction seen for filament and void galaxies. An interpretation of this is that, in filaments and voids, other pathways for $\mathrm{d} \log[Z_g/Z_{\odot}]$ regulation are open. For example, there is more gas untied to galaxies and available in the IGM, such that galaxies that are not undergoing a minor merger are still experiencing a level of cold gas accretion from a pathway that is not functioning in the nodes. When one of these galaxies does experience a minor merger, its $\mathrm{d} \log[Z_g/Z_{\odot}]$ is less-effected than a galaxy in a node. Whilst when a node galaxy experiences a minor merger, this may be its only inflow of lower metallicity gas, meaning its $\mathrm{d} \log[Z_g/Z_{\odot}]$ is far more sensitive to this process. This process would likely only produce significant impacts on the residual of low $f_g$ non-star-forming galaxies, or smaller satellite galaxies.

\subsubsection*{SMBH Growth}

As mentioned, minor mergers demonstrate the largest increases in $\log[Z_{g}/Z_{\sun}]$ early times, suggesting their significant role in the mass evolution of galaxies at this epoch. However, when considering $\mathrm{d} \log[Z_g/Z_{\odot}]$, galaxies with high $\langle \dot{M}_\mathrm{SMBH} \rangle$ show more negative values of $\mathrm{d} \log[Z_g/Z_{\odot}]$ by $0.2$ dex for the node galaxies, compared to an offset relative to the overall MZR of only approximately $0.11$ dex for galaxies undergoing a minor merger. $\langle \dot{M}_\mathrm{SMBH} \rangle$ is a proxy for the AGN activity of a galaxy. We speculate here that this metallicity regulation from SMBH activity might emerge from metal-rich outflows caused by AGN jets (e.g., see the models of \citealt{Moll2007} and the observations of \citealt{Santoro2020}); nevertheless, we would like to caution the readers that our interpretation of HR5 outputs does not provide an unambiguous explanation yet, being based on unconnected models and observations. Fig. \ref{fig:Residual_dMsmbhdt_Evolution} demonstrates this offset relative to the overall MZR compared to galaxies with little to no SMBH growth. Interestingly, the greatest offset of $\mathrm{d} \log[Z_g/Z_{\odot}]$ for filaments and nodes occurs at $z=2.8$, corresponding to the large peak in $\log[\dfrac{\Delta M_{\mathrm{SMBH}}}{\Delta \mathrm{z}}]$ from the right panel of Fig. \ref{fig:Msmbh_dMsmbhdt_Evolution}. The void galaxies experience their maximum offset at $z=2.1$, also corresponding to the large peak in $\log[\dfrac{\Delta M_{\mathrm{SMBH}}}{\Delta \mathrm{z}}]$ for void galaxies in the same figure. This shows a clear difference in how supermassive black holes grow in different environments, especially in the time aspect. It is likely that in dense environments, like nodes and filaments, there is more available matter for the SMBHs to accrete. This means they will grow at an increased rate when compared to the void SMBHs, reaching their peak in $\log[\dfrac{\Delta M_{\mathrm{SMBH}}}{\Delta \mathrm{z}}]$ first. The void SMBHs grow more slowly due to the lower abundances of matter to accrete, and as such, their large peak in activity comes at a later time. 

The values of $\mathrm{d} \log[Z_g/Z_{\odot}]$ of galaxies with high $\langle \dot{M}_\mathrm{SMBH} \rangle$ show a greater offset compared to galaxies with low $\langle \dot{M}_\mathrm{SMBH} \rangle$. This reduction in the offset is of a different magnitude for each environment. Node galaxies are the most sensitive to high SMBH growth with the most negative residuals of $0.25$ dex, filaments take the typical intermediary place, whilst void galaxies show an offset of $0.1$ dex respectively at their individual peaks of SMBH growth. Interestingly, even though the pattern of $\log[\dfrac{\Delta M_{\mathrm{SMBH}}}{\Delta \mathrm{z}}]$ in Fig. \ref{fig:Msmbh_dMsmbhdt_Evolution} is very similar in magnitude and shape for filaments and nodes, the offset in $\mathrm{d} \log[Z_g/Z_{\odot}]$ relative to the overall MZR due to that activity is systematically different. This suggests that the systematic difference in offset from the overall MZR between environments is less due to a difference in SMBH growth, but due to another difference between the galaxies in each environment. Following our speculation above, AGN activity typically causes high metallicity outflows \citep{Santoro2020}, and due to higher density environments containing higher metallicity galaxies \citep{Cooper2008, 2009Ellison, 2017Wu, Andrade2024, Kane2024, Rowntree2024}, these outflows in the nodes would be higher in metallicity than those same outflows in the voids. However, this absolute difference does not explain the more negative $\mathrm{d} \log[Z_g/Z_{\odot}]$ in node galaxies: for these outflows to cause a decrease of $\log[Z_g/Z_{\odot}]$, the ejected gas must be enriched relative to the ambient gas in the galaxy itself. We speculate that the more negative $\mathrm{d} \log[Z_g/Z_{\odot}]$ of node galaxies might therefore reflect their stronger SMBH growth and lower gas fractions, together with the suppression process of their gas replenishment. This picture is supported by simulation studies of AGN outflows. In their analysis of HR5, \citet{Lee2021} note that outflows typically carry metallicities well above the ISM average, and \citet{Choi2020} likewise find AGN-driven winds that are enriched relative to the host ISM ($Z_\mathrm{out}/Z_\mathrm{gal}>1$). We caution, however, that an enriched outflow does not by itself guarantee a downward shift on the MZR: in the chemodynamical simulations of \citet{TaylorKobayashi2015letter} the AGN outflow is instead drawn from metal-poor central cold flows, leaving the galaxy MZR largely unchanged. This reinforces our interpretation that the systematic difference between environments is set primarily by how efficiently each grows its black holes, rather than by outflow metallicity alone.

It is by $z=1.5$ that the magnitude of the residual in $\mathrm{d} \log[Z_g/Z_{\odot}]$ begins to wane as seen in the first panel of Fig. \ref{fig:Residual_dMsmbhdt_Evolution}. There are two smaller peaks in $\log[\dfrac{\Delta M_{\mathrm{SMBH}}}{\Delta \mathrm{z}}]$ that are observed at $z=1.8$ for filament and node galaxies, and at $z=1.2$ for the void galaxies, that are offset by the same redshift as the large peaks. However, due to their magnitude, it is harder to be statistically certain in them and there is no clear offset in $\mathrm{d} \log[Z_g/Z_{\odot}]$ due to them in Fig. \ref{fig:Residual_dMsmbhdt_Evolution}, where we do see reduced $\mathrm{d} \log[Z_g/Z_{\odot}]$ at the particular redshift corresponding to the large peak. Equally, we speculate that this action of gas being pushed out of the galaxy by AGN activity could push the galaxy more rapidly towards quiescence and as such higher metallicities \citep{Peluso2023}, as sufficient time passes, providing another possible, yet not unambiguous, route to explain how higher AGN activity in the nodes and filaments might contribute to, instead, higher metallicities at late times. Remaining in these intermediate times, between $z=1.5$ and $z=3.5$, SMBH growth demonstrates some of the largest impacts on chemical evolution that we observe in this study. In this time-frame SMBH growth is acting to regulate metallicty in all three environments through high metallicity outflows, where node galaxies have more metallicity to lose, and as such show the largest offsets relative to the overall MZR. Node galaxies however are still the most metal rich galaxies observed of all the environments, so even though SMBH growth is impacting their metallicity so strongly, it is not enough to cause them to be metal poor. We speculate that galaxies in the nodes might also be recycling and re-accreting the same enriched gas over their lifetime as it is contained within the cluster (see also \citealt{Tremblay2016, Gaspari2016, Gaspari2017, Fluetsch2019}). A similar recycling of enriched ejecta has been studied in other cosmological simulations, which lends some support to this speculation. \citet{Lee2021} note that a fraction of the metals expelled in HR5 outflows are re-accreted at a later stage, while the FIRE baryon-cycle study of \citet{AnglesAlcazar2017} and the recycled-wind-accretion model of \citet{OppenheimerDave2010} find that re-accreted wind material can supply much of a galaxy's gas over its history, recycling more rapidly around the more massive galaxies that dominate denser environments such as our nodes. Consistent with this, \citet{LinZu2026} show, using DESI and the EAGLE simulation, that continuous accretion of enriched intracluster-medium gas, rather than pristine accretion, dominates the metallicity enhancement of galaxies within the cluster virial radius.

So, although they lose it, and regulation from cold-gas accretion is far less common here, accreted gas from the IGM is already enriched \citep{Peng2014}, leading to a rapid chemical evolution in these environments. On the contrary, AGN activity has also been shown to trigger star formation rate in the galaxy \citep{Kraljic2019}, surrounding galaxies \citep{Fujita2008, Carniani2016, Das2025}, and in the compressed gas at the head of the jet, originally investigated through the positive correlation between AGN luminosity and metallicity \citep{Maiolino2006, Maiolino2019}. This provides a pathway in which, indirectly, faster SMBH growth, which we observe in the nodes, linked to AGN activity and energetic jets, leads to the compression of gas and as such further star formation, which contributes to the chemical evolution of the galaxies by continuing to process gas into heavier metals, \citep{Zubovas2013, VillarMartin2024, Ong2025}. We caution again the readers that most of our interpretations in this discussion section do not provide an unambiguous explanation of HR5 outputs yet, being the discussion based on unconnected models and observations.

It is discussed in \cite{DelPino2023} that AGN gas accretion is inversely proportional to the distance to the central galaxy of a cluster. This is suggested to be an outcome of the low gas content of galaxies in the centers of clusters. This means that when comparing high density environments (nodes and filaments) to low density environments (the void), less negative feedback is experienced by the galaxies within high-density environments, providing another way in which the fundamentals of AGN influence metallicity. 
Once we see that $\mathrm{d} \log[Z_g/Z_{\odot}]$ becomes increasingly less negative, approaching the MZR, from its offsets due to both minor mergers and $\langle \dot{M}_\mathrm{SMBH} \rangle$ at approximately $z=1.5$, we speculate that we enter an era where the chemical evolution of galaxies between the environments is largely dominated by the starvation process, where the stability of a galaxies metallicity and $\mathrm{d} \log[Z_g/Z_{\odot}]$ is determined by the abundance of available pristine gas in the galaxies reservoir, and in the surrounding medium. It is at $z=1$ that we see the $M_\mathrm{gas}$ and $f_g$ of node galaxies drop dramatically, meaning that available pristine gas to accrete has run dry. This is systematically different with density when comparing the three environments. The slowly evolving voids have not experienced an accelerated chemical evolution \citep{Pustilnik2016, Kreckel2016}, and as such show no turnover at this time, becoming the most gas rich population at $z=0.625$. As the intermediate density environment, the filament galaxies do begin to show a turnover by $z=0.625$, however clearly offset from that seen for the nodes. Higher densities in these global environments, through an enhancement physical processes at play, accelerates the consumption of gas and the chemical evolution of galaxies within them. Once this gas runs out, and no more is available to accrete, the strong starvation signal that we see for the nodes begins to show itself. It is interesting that galaxies with very high stellar mass fraction show dramatically increased $\mathrm{d} \log[Z_g/Z_{\odot}]$ compared to those with low stellar mass fraction below $z=1$. In this redshift range, galaxies with high stellar mass fraction show an increase of $0.3$ dex in $\mathrm{d} \log[Z_g/Z_{\odot}]$ compared to galaxies with low stellar mass fraction, this is the largest increase in $\mathrm{d} \log[Z_g/Z_{\odot}]$ we observe when comparing between two populations. This suggests that below $z=1$, these are the galaxies that have successfully converted their gas into stars, which then impart feedback back onto the galaxy through supernovae and stellar winds. With less pristine gas available to regulate metallicity in these dense regions, the evolution of the metallicity begins to runaway, producing the highest metallicities, and highest positive deviations from the MZR. 

With reference to both SFR-based and gas property-based FMRs, our results align with the generalised findings. High $f_g$ is linked with increased SFR. Galaxies with high SFR on the FMR show decreased metallicity and our results also report this same result in absolute $Z_g$ and MZR residual. Low $f_g$ links to the opposite, namely a more starved and low SFR galaxy. On the FMR, this points to increased metallicity across the full stellar mass range, once again agreeing with our results. It is important to note that although the interplay between $Z_g$ and $\mathrm{d} \log[Z_g/Z_{\odot}]$ here suggests they are interchangeable, they are not. One must consider the evolutionary path of two galaxies through the MZR. Assume two galaxies that both start at the same point with low $Z_g$, low $M_\star$ and a positive $\mathrm{d} \log[Z_g/Z_{\odot}]$, sitting just above the overall MZR. The first galaxy has an evolutionary path that has a slope steeper than the MZR, with $M_\star$ increasing concurrently with $Z_g$ as it evolves. Once we look at its properties within the final snapshot it has increased $Z_g$, and a positive $\mathrm{d} \log[Z_g/Z_{\odot}]$. In the case of the other galaxy, its evolutionary path has a slope that is shallower than that of the overall MZR. Once again, as this galaxy evolves, $M_\star$ increases concurrently with $Z_g$, however at the final snapshot, this galaxy now has a negative residual, as the average $Z_g$ of the MZR also increases as the galaxy evolves. This shows the case in which an increased $Z_g$ and a reduced, negative $\mathrm{d} \log[Z_g/Z_{\odot}]$, can co-exist. The galaxy does not require a drop in $Z_g$ for a reduced or negative residual, only to evolve slower than the MZR itself.

These results also suggest that the formation time of galaxies is a key environmentally-driven factor in the overall distribution of galaxies in the universe, and throughout the cosmic web. The voids are regions where star formation, relative to the overall star formation history of the Universe, is delayed, whilst the more dense environments are those where star formation is accelerated. Older galaxies are therefore seeded in higher density environments under-going the phases of their lives whilst the galaxies born in lower densities lag behind. Metallicity is a residue of the unique star formation history in each environment, which in itself is governed by the alternate physical processes, and their efficiencies, that are at play, which overall gives rise to the patterns that we observe in this study.


\section{Conclusion}
\label{sec:Conclusions}

We have presented an analysis of gas metallicty, $Z_g$, and the stellar mass-gas metallicity relation, MZR, in the context of three main cosmic environments, nodes, filaments and voids across 86 snapshots in the Horizon Run 5 (HR5) simulation between $z=0.625$ and $z=5$. These environments were defined by a careful calibration of the subhalo catalogues and the {\tt T-ReX} filament finder. By fitting the MZR with a linear regression at these 86 snapshots, limited only to the linear region of the MZR before the turnover begins, $\log( M_{\star}/\text{M}_{\odot} ) \lesssim 10.7$, we defined the MZR residual, $\mathrm{d} \log[Z_g/Z_{\odot}]$, which measures the scatter in the MZR at each specific moment in time. 
Within this space, we produced results regarding the influence of three key physical processes, major and minor galaxy mergers, SMBH growth, and gas gain and loss, on the chemical evolution of galaxies across time.

\begin{enumerate}

\item Galaxies that end up in nodes at $z=0.625$ build up the majority of their high $\mathrm{d} \log[Z_g/Z_{\odot}]$ below $z=1.5$, at later times. This period of time also lines up with when the node galaxies experience a dramatic drop in $M_\mathrm{gas}$. This suggests that there is no more available pristine gas in the node environments to accrete that would replenish their gas reservoirs. This would regulate the $Z_{g}$ of these galaxies, however this process has halted, or only more metal rich gas can inflow, leading to a dramatic increase in $\mathrm{d} \log[Z_g/Z_{\odot}]$. 

\item Galaxies that have high $f_{g}$, across all redshifts, show negative residuals, whilst galaxies with low $f_{g}$ show positive residuals. Below $z=1.5$, galaxies with low $f_{g}$ demonstrate the highest residual values we observe at $z \approx0.8$ with $\delta \log[Z_g/Z_{\odot}] = 0.2$.This further suggests the importance of starvation and the lack of gas accretion in highly dense environments as a driver of the scatter in the MZR at late times. 

\item Galaxies in all three environments that are experiencing minor mergers show a dramatic $0.23$ dex increase in $\log[{Z_g}]$ at the earliest snapshot, $z\approx5$. This difference reduces as redshift decreases down to $0$ dex for all environments by $z=2$, and no significant difference mergers towards lower redshifts. Across all redshifts, galaxies experiencing minor mergers show reduced $\mathrm{d} \log[Z_g/Z_{\odot}]$ compared to galaxies without minor mergers in the same environment, with the largest decreases seen in the node galaxies of $0.11$ dex.

These together suggest that large galaxies accreting smaller ones, through minor mergers at early times, lead to higher mass galaxies that can use the available gas from this process to evolve to higher metallicities, producing the MZR. Due to the accretion of gas from the minor mergers, their $\mathrm{d} \log[Z_g/Z_{\odot}]$, in the short term, is regulated but this gas can then lead to star formation. This effect is strongest at early times when low mass galaxies are most abundant. Minor mergers are most abundant in nodes at all redshifts, followed by the filaments, then the voids.

\item Galaxies that are experiencing major mergers show minimal difference in their evolutions of $\log[Z_g]$. These galaxies also show minimal differences in $\mathrm{d} \log[Z_g/Z_{\odot}]$ across all redshifts, the largest being an approximate $0.06$ dex difference between majorly merging and non merging node galaxies at later times. This suggests overall that major mergers have little effect on chemical evolution, and the effect they do have reduces the $\mathrm{d} \log[Z_g/Z_{\odot}]$ of galaxies in dense environments at late times.

\item SMBH growth within galaxies in HR5 demonstrates two peaks. The first large peak occurs for the nodes and filaments at $z=2.8$, corresponding to $\log(M_\mathrm{SMBH}/M_\odot)\approx4.3 $, this large peak then occurs for voids at a later time of $z=2.1$, corresponding to the same mass threshold. A second smaller peak is then seen for filaments and voids at $z=1.9$, corresponding to $\log(M_\mathrm{SMBH}/M_\odot)\approx4.5 $. The second void peak corresponding to the same mass threshold then occurs again later at $z=1.1$. This shows that the void populations' BHs grow at a slower rate than those in higher-density environments like filaments and voids. Exhibiting lower SMBH growth across all redshifts, with a lower first peak. This SMBH growth peaks at intermediate times, between $z=1$ and $z=3.5$.

\item Galaxies with high SMBH growth in all environments at intermediate redshifts, between $z=1$ and $z=3.5$, show negative residuals. This effect is strongest for the nodes, then filaments, and weakest for the voids. This includes the most negative $\mathrm{d} \log[Z_g/Z_{\odot}]$ we see of $0.25$ dex for the nodes at $z=2.7$. This shows that SMBH growth is responsible for driving negative $\mathrm{d} \log[Z_g/Z_{\odot}]$ values at intermediate times, likely effecting node and filament galaxies more due to their higher average SMBH growth.

\item Galaxies with low SMBH growth show similar magnitude of positive residuals to the total average population. However, at late times, all environments show no significant offset in $\mathrm{d}\log[Z_g/Z_{\odot}]$ relative to the overall MZR. There is also a general trend in this low SMBH growth population for $\mathrm{d} \log[Z_g/Z_{\odot}]$ to decrease with time in filaments and voids, however the trend is flat for the nodes.

\end{enumerate}

Overall, at early times, minor mergers drive chemical evolution by the highest magnitude, influencing the nodes the most due to the higher abundance of minor mergers in this environment across all redshifts. The filaments and voids are less effected by these minor mergers in the $\mathrm{d} \log[Z_g/Z_{\odot}]$ vs redshift space, whilst the nodes show a consistently reduced residual across all of time, showing that minor mergers do also drive the MZR scatter significantly at all redshifts. As we move to intermediate redshifts, SMBH growth takes more prominence, leading to dramatic reductions in $\mathrm{d} \log[Z_g/Z_{\odot}]$ of $-0.1$ dex for the void population, $-0.2$ dex for the filament population, and a peak of $-0.25$ dex for the node population. SMBH growth is highest in these intermediate times, likely leading to the regulation of metallicity due to enriched outflows triggered by the SMBH growth. SMBH growth is higher in nodes, and more dense environments, explaining why we see the largest drops $\mathrm{d} \log[Z_g/Z_{\odot}]$ for these environments. $\mathrm{d} \log[Z_g/Z_{\odot}]$ then recovers as time passes, increasing once again as SMBH growth wanes. At late times, below $z=1.5$, we then observe the effect of strangulation take precedence in the high density environments, as the gas reservoirs of these galaxies run dry, and there is no more available gas in these high density environments to accrete. It is within this redshift period, that we, not only, see a dramatic rise in residual for node galaxies with low gas mass, with an increase of $0.2$ dex leading to our highest reported values of $\mathrm{d} \log[Z_g/Z_{\odot}]$, but also a dramatic drop in $M_\mathrm{gas}$ of galaxies in this environment.

These results help to disentangle the effects of these key physical processes in the context of unique cosmic environments. This study shows us when and where these processes take precedence, driving the MZR and chemical evolution by different amounts at different points in time and space. The combination of the results together helps to paint a story of how galaxies in different environments experience different evolutionary pathways based on their environment, helping to understand just how galaxies of all shapes and sizes exist at present day. 
We are now in a time in which observations can provide insight into these questions. Subaru's Prime Focus Spectrograph (PFS) \citep{Tamura2016} already provided some useful early data needed to bring these ideas into an observational space. DESI's early data release \citep{DESI2016, DESI2024} currently provides powerful measurements of spectra and large-scale structure measurements simultaneously. EUCLID \cite{Euclid2022} plans to have its first data release on June 24th 2026, providing more of the same large-sample size, high quality spectra. Finally, with the launch of the PRIMA mission \citep{Glenn2025}, as early as 2031, the results presented in this paper will truly be testable from multiple surveys based in our own Universe. We hope that this study can help to underpin and provide comparison for these future fundamental works.

\section*{Acknowledgements}

We thank the reviewer for their detailed review of the manuscript. This review significantly improved the quality of the work, and helped to ground it within its surrounding fields. We acknowledge the support of STFC (through the University of Hull's Consolidated Grant ST/R000840/1) and ongoing access to {\tt viper}, the University of Hull High-Performance Computing Facility. FV thanks the National Science Foundation (NSF, USA) under grant No. PHY-1430152 (JINA Center for the Evolution of the Elements). We acknowledge the support of computing resources at the Center for Advanced Computation (CAC) at KIAS. AS was supported by KIAS Individual Grants (PG080901) and acknowledges support from the UK Research and Innovation (UKRI) Nottingham Astronomy consolidated grant ST/X000982/1, Astronomy and Astrophysics
at the University of Nottingham- 2023 to 2026 (PI: Simon Dye). This work is partially supported by the grant GALBAR ANR-25-CE31-4684 and from the CNRS through the MITI interdisciplinary programs.

\section*{Data Availability Statement}

The data underlying this work is able to be shared upon reasonable request.

\appendix

\bibliographystyle{mnras}
\bibliography{references}

\end{document}